\documentclass[aps,pra,twocolumn,amsmath,amssymb,longbibliography,10pt]{revtex4-1}
\usepackage{amsfonts}
\usepackage{amsmath}
\usepackage{amssymb}
\usepackage{graphicx}
\usepackage[T1]{fontenc}
\usepackage[utf8]{inputenc}
\usepackage{hyperref}
\usepackage[dvipsnames]{xcolor}
\usepackage{color}
\usepackage{bm}
\usepackage{comment}
\usepackage{mathtools}
\usepackage[section]{placeins}
\usepackage[capitalize]{cleveref}
\usepackage{footnote}
\usepackage{etoolbox}
\usepackage{ulem}
\usepackage{pgfplots} 
\usetikzlibrary{plotmarks}
\pgfplotsset{compat=newest} 
\pgfplotsset{plot coordinates/math parser=false}

\crefrangelabelformat{equation}{(#3#1#4--#5\crefstripprefix{#1}{#2}#6)}

\begin{document}
\title{Screening and the Pinch Point Paradox in Spin Ice}

\author{Mikael Twengstr{\"o}m} 
\affiliation{Department of Physics, Royal Institute of Technology, SE-106 91 Stockholm, Sweden}
\author{Patrik Henelius}
\affiliation{Department of Physics, Royal Institute of Technology, SE-106 91 Stockholm, Sweden}
\affiliation{Faculty of Science and Engineering,  \r{A}bo Akademi University, \r{A}bo, Finland}
\author{Steven T. Bramwell}
\affiliation{London Centre for Nanotechnology and Department of Physics and Astronomy, University College London, 17-19 Gordon Street, London, WC1H OAH, U.K.}

\begin{abstract}  
A pinch point singularity in the structure factor characterizes an important class of condensed matter that is a counterpoint to the paradigm of broken symmetry. This class includes water ice, charge ice and spin ice. Of these, dipolar spin ice affords the the pre-eminent model system because it has a well-established Hamiltonian, is simple enough to allow analytical theory and numerical simulation, and is well represented in experiment by Dy$_{2}$Ti$_{2}$O$_{7}$ and Ho$_{2}$Ti$_{2}$O$_{7}$. Nevertheless it is a considerable challenge to resolve the pinch points in simulation or experiment as they represent a very long range correlation. Here we present very high resolution simulations of the polarized neutron scattering structure factor of dipolar spin ice and new analytical theory of the pinch point profiles. We compare these with existing theory and experiment. We find that our simulations are consistent with theories that reveal the pinch points to be infinitely sharp, as a result of unscreened dipolar fields. However, neither simulation nor these theories are consistent with experiments, which instead is quantitatively captured by a theory that allows for screening of the dipolar fields and consequent strong broadening of the pinch points.  This striking paradox is not easily resolved: broadening of the pinch points by random disorder seems to have been ruled out by existing theory, while deficiencies in the Hamiltonian description are not relevant. Intriguingly, we are left to consider the role of quantum fluctuations or the possibility of a fundamental correction to either the standard method of simulating dipolar systems, or the theory of polarized neutron scattering. More generally, our results may have relevance far beyond ice systems. For example, spin ice is a model Debye-H\"uckel (magnetic) electrolyte, so our basic observation that the screening length may diverge while the Debye length remains constant, may hint at a solution to the topical problem of `underscreening' in concentrated ionic liquids.   
\end{abstract}

\maketitle
\section{Introduction}

The decomposition of the scattering amplitude into the product of form factor and structure factor is a key device in the theory of scattering from condensed matter. While the form factor represents the scattering amplitude of an object, the structure factor represents the correlation between objects. Arrays of dissimilar objects may exhibit similar structure factors because their correlations are similar. Hence the classification of structure factors plays a key role in classifying the types of physical state that may occur in condensed matter systems and beyond.  

\begin{figure}[!htb]
		\centering
  		\resizebox{\hsize}{!}{\includegraphics{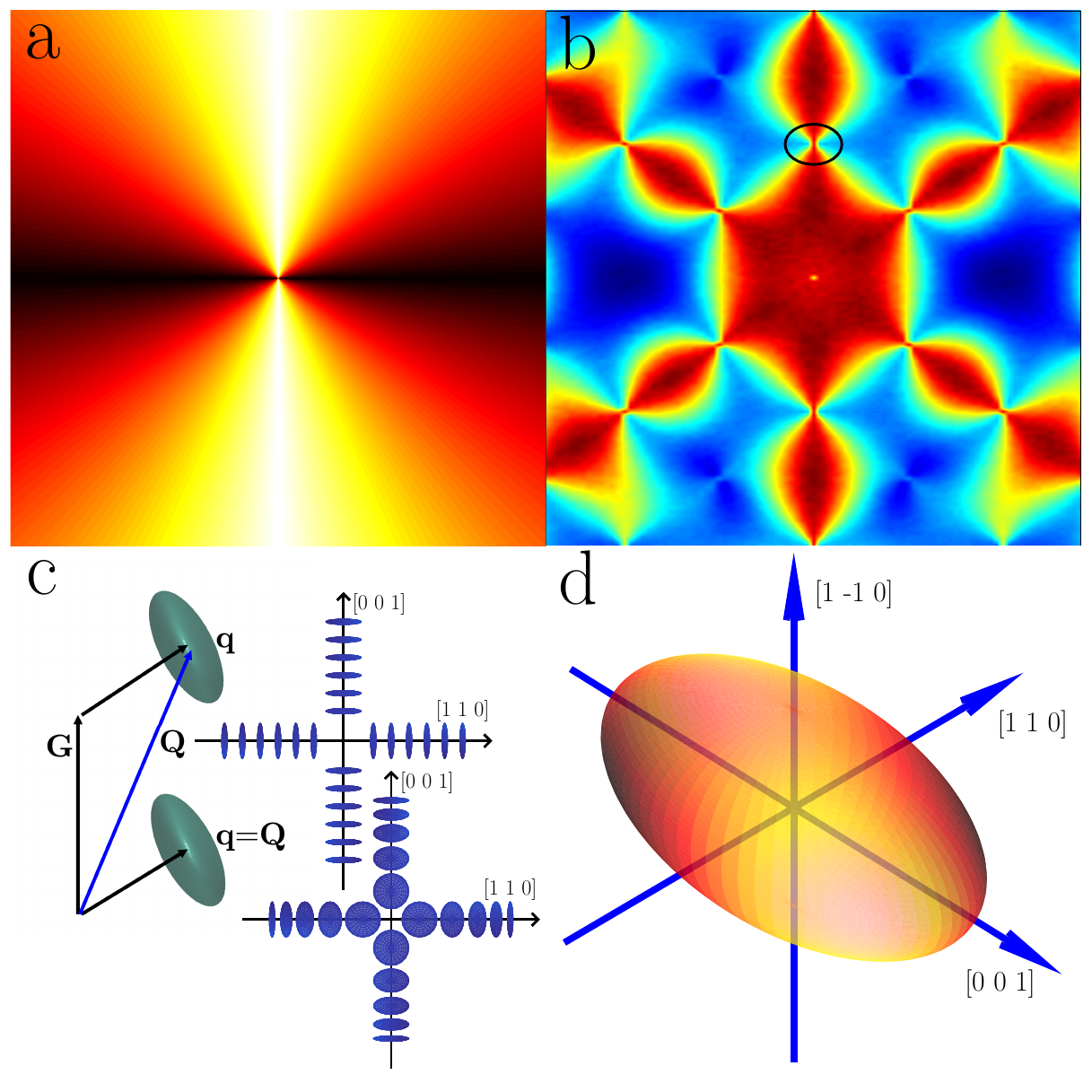}}
 		 \caption{(a) A pinch point. The function approaches a singularity in which its value at $q$ = 0 (centre of plot) depends on the orientation of the any line passing through $q = 0$. (b) The (simulated) spin flip (SF) pattern of spin ice in the $1\bar{1}0$ scattering plane of reciprocal space: the pinch point at $002$ is circled. (c) Left: relation of the scattering vector ${\bf Q}$, reciprocal lattice vector ${\bf G}$ and ${\bf q} = {\bf Q}-{\bf G}$, illustrating the representation ellipsoid of the structure factor tensor ${S^{\alpha \beta}}({\bf q})$. The ellipsoid has different transverse projections on ${\bf Q}$ in the first zone (where ${\bf Q} = {\bf q}$) and higher zones (where ${\bf Q = G+q }$): it is these projections that are seen in neutron scattering. Right: how the ellipsoid varies with ${\bf q}$ in the case of sharp pinch points (upper, where the ellipsoid is eccentric at all finite $q$) and broadened pinch points (lower, where it becomes a sphere for small $q$). (d) Representation ellipsoid and principal axes of the inverse structure factor tensor along the line $hh2$ (scan `across' the pinch point at $002$), as discussed in this work. }
 		 \label{figure:1}
\end{figure}

The various classes of structure factor $S^{\alpha\beta}({\bf q})$ may be elucidated by considering possible singularities in the function. The pinch point~\cref{figure:1}\,a is a type of singularity that is expected, and may be observed, in at least two types of system:  firstly, dipolar systems, such as ferromagnets~\cite{Krivoglaz,Als}, and  secondly, ice-rule systems, including hydrogen bonded ferroelectrics~\cite{Paul,Havlin,YA,YAM}, water ice~\cite{Nield, Wehinger}, spin ice~\cite{MS,Fennell_kag,Fennell,Chang,Sen,RRB}, ionic ice~\cite{Anderson,Fennell2019}, artificial spin ice~\cite{Perrin,Ostman,Farhan}, quantum spin ice~\cite{Benton,Fennell_TTO, Sibille} and antiferromagnetic spin liquids~\cite{Zinkin,Conlon,Ballou,Canals,Henley}. Such systems have the notable feature that they enter a highly correlated low-temperature state without any symmetry breaking. As such, they provide a valuable contrast to the usual Landau paradigm of condensed matter, where formation of low temperature states is invariably accompanied by symmetry breaking. 

The pinch point structure factor is the principal characteristic and diagnostic of such states (in some, but not all, of the above examples, Pauling entropy~\cite{Pauling, Ramirez, Giblin} is another key characteristic).  However, despite the paradigm nature and widespread relevance of the phenomenon, there have been very few direct confrontations of theory and experiment for the pinch point structure factors. The reason seems to be that any approach to the singularity is difficult to achieve in both numerical simulation and experiment: the correlation that gives rise to the pinch point is very subtle and long ranged, meaning that the amplitude of $S^{\alpha\beta}({\bf q})$ is everywhere very small and high resolution is required to distinguish its fine features as ${\bf q\rightarrow 0}$. In addition, well-defined model systems are very scarce. Spin ice, whose pinch points arise from the combined effects of dipolar and ice-rule correlations~\cite{Bramwell_Gingras}, stands out as the exception to this particular rule: it is a nearly ideal model system which lends itself well to advanced experiment and numerical modelling.  

\begin{figure}[!htb]
		\centering
  		\resizebox{\hsize}{!}{\includegraphics{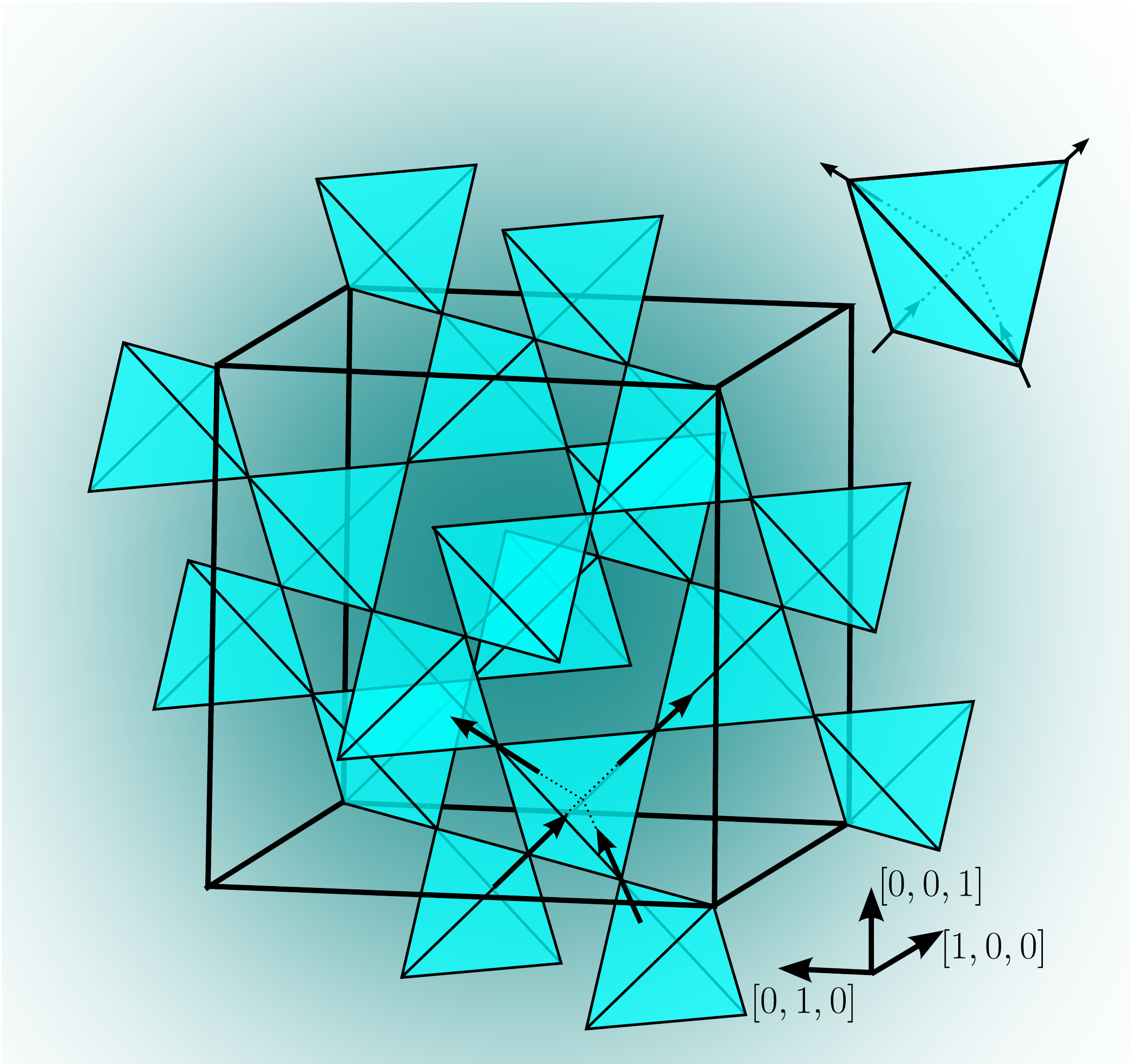}}
 		 \caption{The conventional cubic unit cell of the pyrochlore lattice consisting of $16$ lattice sites (corners of the tetrahedra are the sites). The spins reside on the corners of the tetrahedra and their respective lattice vectors are pinned to the local $\left\langle111\right\rangle$-directions due to the nature of the crystal fields. A spin thus points towards the center of one tetrahedron whilst pointing away from the center of another and vice versa. Inset (upper right): The ground state of spin ice consists of a spin configuration for which each tetrahedron exhibits two spins pointing in and two spins pointing out.}
 		 \label{figure:2}
\end{figure}

The idea of spin ice (see~\cref{figure:2}) was invented~\cite{Harris,BramwellHarris} largely to describe the unusual correlations and absence of long range order observed in neutron scattering experiments on ${\rm Ho_2Ti_2O_7}$. Subsequently, a very deep and consistent understanding of the many interesting properties of spin ice has been developed in terms of the microscopic dipolar spin ice (DSI) model~\cite{DenHertog}, with spin Hamiltonian consisting of exchange and dipolar terms: 
\begin{align}\label{DSI}
    \mathcal{H}=\mathcal{H}_{\rm exchange}+ \mathcal{H}_{\rm dipolar}. 
\end{align}
A truncation of the dipole-dipole interaction to near neighbor in~\cref{DSI} gives a near neighbor ferromagnetic coupling and the original near--neighbor spin ice (NNSI) model of Refs.~\cite{Harris,BramwellHarris}, which remains a valuable point of reference. Today, DSI (along with its extensions~\cite{Yavorskii}) is fully accepted as the basic microscopic description of spin ice, one that leads eventually to some surprising properties, including its residual (Pauling) entropy~\cite{Ramirez}, emergent electromagnetism~\cite{Ryzhkin,CMS}, fractionalization~\cite{CMS} and fragmentation~\cite{BrooksBartlett,Petit}. Yet, despite this progress, the neutron scattering structure factor that first inspired the discovery of spin ice, is not yet fully understood, as we discuss in this paper. In particular, we find that the pinch point profile, that summarizes the long ranged correlations in spin ice, presents a clear contradiction between theory and experiment. 

Motivations for studying the pinch points in classical spin ices like Ho$_{2}$Ti$_{2}$O$_{7}$ and Dy$_{2}$Ti$_{2}$O$_{7}$ include not only their above-mentioned status as model representatives of a large class of condensed matter, but also their more recently-established status as model Coulomb fluids~\cite{CMS,CMSDH,Kaiser,KaiserPRL,Bovo,Paulsen}. Here they afford perhaps unique examples of a model lattice Coulomb fluid (of magnetic monopoles) that quantitatively obeys Debye-H\"uckel theory in both the dilute and concentrated regimes~\cite{Kaiser}. Neutron scattering affords a rare opportunity to directly image the field correlations and screening in such a model experimental Coulomb fluid. To do this may shed some light on foundational issues of the theory of Coulomb fluids, for example the highly topical problem of `underscreening' in concentrated ionic liquids, where the experimentally measured screening length appears to diverge, while the Debye length remains finite~\cite{Lee2015,Lee}. 

\medskip
\noindent
{\bf Structure factor of spin ice --}
A complete characterization of the static correlations in spin ice can only be achieved using polarized neutron scattering~\cite{Fennell}. The reason is that the dipole-dipole interaction breaks the rotational invariance of the correlation function, such that the structure factor tensor $S^{\alpha \beta}({\bf q})$ is characterized by two numbers, corresponding to the longitudinal (to ${\bf q}$) and transverse eigenvalues, $S^{\rm L}$ and $S^{\rm T}$ respectively. Polarized neutrons are required to separate these numbers, which in unpolarized neutron scattering appear only in combination.

Here we define ${\bf Q} = {\bf q} + {\bf G} $, where ${\bf Q}$ is the scattering vector and ${\bf G}$ is a reciprocal lattice vector (\cref{figure:1}c). The neutron scattering differential cross section measures the transverse projection of the structure factor on the scattering vector~\cite{Marshall-Lowde}:
\begin{equation}
\frac{\textup{d}\sigma^{\alpha\beta}}{\textup{d}\Omega} = A \left(\delta_{\alpha \beta}- \hat{Q}_\alpha\hat{Q}_\beta\right)\, S^{\alpha\beta}({\bf Q}), 
\end{equation}
(we set $A=1$ henceforth).  Within an unimportant caveat discussed below, the structure factor is the same in each Brillouin zone (hence characterized as $S^{\alpha \beta}({\bf q})$), but its projection may vary from zone to zone, as seen in~\cref{figure:1}c.   

The origin of the pinch points may be visualized by representing the inverse $S^{\alpha \beta}({\bf q})$ tensor as an ellipsoid of dimensions $\sqrt{S^{\rm L}({\bf q})}\times \left(\sqrt{S^{\rm T}({\bf q})}\right)^2$,  oriented with principal (L) axis parallel to ${\bf q}$~\cite{Bramwell_NComms, Nye}.  In a dipolar ferromagnet (say) the ellipsoid is eccentric (i.e. $S^{\rm L}\ne S^{\rm T}$) only at small $q$~\cite{Als}, but in spin ice, the conspiracy of ice rules and very strong dipolar forces impose a non-spherical structure factor tensor over nearly the entire Brillouin zone.~\cite{Bramwell_NComms}. Considering the in plane projection of this ellipsoid on ${\bf Q}$ (\cref{figure:1}c), the pattern shown in~\cref{figure:1}b is easily inferred. If the ellipsoid remains eccentric as $q\rightarrow 0^+$ then the pinch points are infinitely sharp; if it evolves to a sphere in that limit they are broadened (\cref{figure:1}c). Precisely at $q=0$, in either case, the structure factor ellipsoid is a sphere, in accord with the crystal symmetry.   

We focus on the case of ${\bf G} = 002$ in the $1\bar{1}0$ scattering plane used in the experiments~\cite{Fennell, Chang}. If we inspect wavevectors ${\bf Q} = hh(l=2)$ within the $002$ zone, then $h\bar{h}0$ and $hh0$ define two principal axes of the tensor (\cref{figure:1}d). Non-spin flip (NSF) scattering then measures the out of plane, transverse (to ${\bf q}$) eigenvalue, while spin flip (SF) scattering measures a mixture of longitudinal and transverse eigenvalues in plane: with the projection perpendicular to ${\bf Q}$, this mixture creates the pinch points in the SF channel (\cref{figure:1}b), while the NSF scattering, as a pure eigenvalue, remains a periodic function.  However, it turns out (see below) that a SF scan along $hh2$ picks out only the longitudinal eigenvalue up to about $h = 0.5$, so this particular scan `across' the pinch point isolates the two eigenvalues needed to characterize the correlations in spin ice: $S^{\rm L}$ in the SF channel and $S^{\rm T}$ in the 
NSF channel. Thus, except very near the zone boundary, the entire correlation function of spin ice is contained in the two channels (SF,NSF) of the single line scan across the $002$ zone centre in the $1\bar{1}0$ scattering plane.

The transverse and longitudinal structure factors represent the correlations of the Helmholtz decomposed components of the magnetization~\cite{Bramwell_NComms}. In the low temperature limit the inverse structure factor tensor becomes a disc ($S^{\rm L} = 0 $), reflecting a divergence free state, the so-called Coulomb phase~\cite{Henley}. In this special case there is a singularity in the correlation function as one tensor eigenvalue, $S^{\rm L}$, disappears. The more general case of an oblate spheroid that is invariant with ${\bf q}$ has been termed the `harmonic phase'~\cite{Bramwell_NComms} as it is implied by Onsager's theory~\cite{Onsager} of dipolar fluids, which contain only harmonic (divergence and curl free) fields. In this context, we ask the question, what happens at finite temperature? As $q\rightarrow 0^+$, does the structure factor tensor of spin ice becomes a sphere (hence broadened pinch points) or does it become an oblate spheroid (hence infinitely sharp pinch points)? These are hard questions to answer as the long range correlations that give rise to the pinch points make meaningful simulations and experiments quite challenging.  Nevertheless, we discuss these questions with respect to new, high resolution, numerical simulations of NNSI and DSI, as well as with respect to existing experimental data~\cite{Fennell, Chang} and analytical theories~\cite{Sen, RRB}. 

For completeness we note the subtle caveat mentioned above -- that happily proves to be unimportant. This arises from the glide symmetry that emerges in the Fd$\bar{3}$m space group of spin ice. With respect to the face centred cubic reciprocal lattice $\{{\bf G}\}$ the function $S^{\alpha\beta}{(\bf Q})$ is modulated, such that unit cell in reciprocal space is doubled. This means that every second zone is inequivalent: for example the function is different in the zones defined by ${\bf G} = 002, 004$. We discuss this point in detail elsewhere~\cite{ZoneCenterPaper} where we refer to zones like $002$ as `false zones' and zones like $004$ as `true' zones. The fact that there are two types of zone is the only reason that experiment can clearly resolve the pinch points in spin ice: at the true zone centres the nuclear Bragg peak has its full intensity and this will always obscure the subtle magnetic diffuse scattering, even in a polarized experiment. In contrast, at a false zone centre (e.g. $002$), the nuclear intensity is zero, making the magnetic diffuse scattering accessible to precise measurement. However this advantage potentially comes at a price: at a false centre $S^{\alpha \beta}({\bf Q})$ is not necessarily a measure of magnetization fluctuations, the quantity of most immediate interest. Yet there is a fortunate crystallographic coincidence that renders this point irrelevant to the present study: scans across the pinch point at $002$ (surprisingly) do yield the magnetization fluctuations as required (see Section III). 

\medskip
{\bf Plan of the paper --}
The paper is organized as follows. In section II we describe details of our numerical simulations. In Section III we present an analytical theory of the correlations that allows for different scenarios --  particularly as regards screening of the dipolar fields -- and we compare these with our simulations and other analytical theory~\cite{Sen, RRB}. We find that this theory accounts very accurately for the simulated pinch point profiles in the absence of field screening. In section IV we compare these results with experiment~\cite{Fennell,Chang} and expose a very clear disagreement between experiment and simulation.  However, we find that we can capture the experimental data by allowing screening of dipolar fields in the theory. This defines the pinch point paradox. In the final section, V, we summarize and discuss our main results, including possible resolutions of the paradox.  

\section{Numerical Simulations}
In the spin ice materials ${\rm R_2Ti_2O_7}$, rare earth ions R = Ho or Dy are located on the points of a cubic pyrochlore lattice of corner-sharing tetrahedra~\cite{Pyrochlore} (see~\cref{figure:2}). The trigonal crystal field enforces a doublet ground state for each ion~\cite{crystal1,crystal2,crystal3,Rau_Gingras_2015} and establishes a local Ising-like~\cite{Rau_Gingras_2015} confinement with effective two-state spins ${\bf S}_i$ pointing between the centers of each pair of adjacent tetrahedra. The associated magnetic moments are very large, $\mu \approx 10~\mu_{\rm B}$, with the consequence that dipole-dipole interactions are particularly strong in these materials. For modelling purposes we consider classical spins of unit length. The dipolar spin ice model (DSM) combines the long-ranged dipolar interaction with short-ranged exchange terms~\cite{DenHertog}: 
\begin{equation}
\mathcal{H}=J_1\sum_{\langle i,j\rangle }\mathbf{S}_i\cdot\mathbf{S}_j+Da^3\sum_{i>j}\frac{\mathbf{S}_{i}\cdot\mathbf{S}_{j}-3\left(\mathbf{\hat{r}}_{ij}\cdot\mathbf{S}_i\right)\left(\mathbf{\hat{r}}_{ij}\cdot\mathbf{S}_j\right)}{r^{3}_{ij}}.\label{sDSMmodel}
\end{equation}
Here $D$ is the dipolar interaction constant, $r_{ij}$ the distance between spin $i$ and $j$ and $J_1$ the nearest-neighbor exchange interaction~\cite{Bramwell_Gingras}.
Considering only nearest-neighbor interactions, the model reduces to the nearest-neighbor spin ice model (NNSI), which gives an accurate approximation to DSI down to about 0.6 K~\cite{melko04}. The NNSI model has effective ferromagnetic exchange parameter $J_{\rm eff}$, which takes the values 1.1 K for ${\rm Dy_2Ti_2O_7}$ and 1.9 K for ${\rm Ho_2Ti_2O_7}$~\cite{Bramwell_Gingras}. The NNSI model maps 
exactly~\cite{BramwellHarris} to Pauling's model of (cubic) water ice, hence exhibiting a degenerate ground state with strong correlations, but no long range order (see~\cref{figure:2}). By introducing the dipolar interaction, the degeneracy of the NNSI model is weakly broken and an ordering transition at very low temperature is induced. However, this is not relevant at the temperatures ($T > 1$ K) considered in this paper, where DSI behaves qualitatively like NNSI. In addition to the nearest-neighbor exchange interactions $J_1$, the generalized spin ice model (g{--}DSM) contains second and third nearest-neighbor interactions $J_2, J_{3a}$ and $J_{3b}$.  A set of parameter values were previously determined for ${\rm Dy_2Ti_2O_7}$ ($J_1=3.41$ K, $J_2=-0.14$ K, $J_{3a}=J_{3b}=0.025$ K) which models unpolarized neutron scattering and bulk thermodynamic properties at a quantitative level~\cite{Yavorskii}.

We previously reported that demagnetizing effects are a pure outcome of dipolar interactions in highly correlated systems~\cite{prm_twengstrom_2017}, i.e. the exchange interactions do not alter the shape-dependent physics at the zone centre. Nevertheless the diffuse scattering elsewhere in the Brillouin zone may depend upon exchange, as discussed subsequently. To model the neutron structure factor in more detail we reconstruct the Fourier transform of the spin-spin correlation function in $\mathbf{Q}$-space. Following the use of a parallel Monte Carlo code~\cite{Giblin} that exploits the symmetry of the dipolar interactions~\cite{ewald21}, we use periodic Ewald boundary conditions~\cite{ewald21} and a loop algorithm~\cite{Melko} to speed up equilibration when needed at low temperature. 

The surface dependence is treated via spherical boundary conditions by considering the addition of a microscopic term to the Ewald sum~\cite{ZoneCenterPaper,mtthesis}. The choice of this boundary term is the foundation of the conditional convergence of the dipolar sum and exploits a way to manage the demagnetizing effects.

High resolution is key to this study. The allowed $\mathbf{Q}$-points were determined by considering only the set of points in which the discrete Fourier transform is defined, i.e. the inverse cubical system size, $1/L$. For example, the high resolution of $L=16$ which we reach corresponds to $65536$ particles and a resolution of $1/L=0.0625$ in units of $2\pi$. Considering that the dipolar systems under study occupy an ordered lattice, parallel computing is essential, given that a system of $65536$ particles corresponds to almost 4.3 billion interactions per Monte Carlo step. Note that we do not employ any interpolation schemes such as the Nyquist--Shannon sampling theorem as this would defeat our purpose of establishing the actual line shapes and their comparison with experiment.

\section{Analytical theory of \texorpdfstring{$S^{\alpha \beta}({\bf Q})$}{Sab(Q)}}

In this section we derive closed-form analytical expressions for the eigenvalues of the structure factor tensor $S^{\alpha \beta}({\bf Q})$ and compare them with our simulations. For dipolar spin ice,  
the scattering functions we derive turn out to be consistent with those of Refs.~\cite{Sen,RRB} but in addition, we have quantified all parameters and controlled boundary effects. Further, our derivation makes a connection with Ref.~\cite{Bramwell_NComms} where the structure factor tensor of spin ice was derived from a very different viewpoint: that of the Onsager dipolar fluid~\cite{Onsager}. 

The theory presented below assumes a continuum magnetization ${\bf M}({\bf r})$. This approximation can only really be valid with respect to magnetization defined on a Bravais lattice, such that a primitive, unit cell can be defined, which, on repetition just fills all space. The pyrochlore lattice (\cref{figure:2}) is a face centred cubic ($F$) Bravais lattice with a four point tetrahedral basis. Hence ${\bf M}$ should be interpreted as the magnetization  averaged over the four $\langle 111\rangle$ spins of the spin ice basis: 
\begin{align}
{\bf M} = g\mu_{\rm B} v^{-1}\sum_{i=1}^4 {\bf s}_{i},
\end{align}
where $g$ is the g-factor, $v$ is the primitive unit cell volume and the  spins
\begin{align}
{\bf s}_1 &= (1/\sqrt{3})(\phantom{-}1,\phantom{-}1,\phantom{-}1),\notag\\ 
{\bf s}_2 &= (1/\sqrt{3})(-1,-1,\phantom{-}1),\notag\\ 
{\bf s}_3 &= (1/\sqrt{3})(\phantom{-}1,-1,-1),\notag\\ 
{\bf s}_4 &= (1/\sqrt{3})(-1,\phantom{-}1,-1),
\end{align}
are located at
\begin{align}
{\bf r}_1 &= (\phantom{/4}0,\phantom{/4}0,\phantom{/4}0),\notag\\ 
{\bf r}_2 &= (1/4,1/4,\phantom{/4}0),\notag\\
{\bf r}_3 &= (\phantom{/4}0,1/4,1/4),\notag\\ 
{\bf r}_4 &= (1/4,\phantom{/4}0,1/4),
\end{align}
respectively in the conventional cubic unit cell. It is straightforward to demonstrate that with respect to the axes $x = [1\bar{1}0]$, $y = [110]$, $z= [001]$, the structure factors probed in the NSF and SF channels,  $S^{xx}$ and $S^{yy}$, measure the  fluctuations $\langle M^x({\bf q}) M^x(-{\bf q})\rangle$ and $\langle M^y({\bf q}) M^y(-{\bf q})\rangle$. Because $002$ is a `false' zone centre (see above) this is not true of the third $(zz)$ component. The difference arises because $zz$ structure factor samples fluctuations of four spins per tetrahedron in the combination $\sqrt{1/3}(s_1^z+s_2^z-s_3^z-s_4^z)$, which is not the same as $M^{z}$, while the $xx,yy$ structure factors sample two spin combinations e.g. $\sqrt{2/3} (s_1^x+ s_2^x)$ and $\sqrt{2/3} (s_3^y+ s_4^y)$, which are the same as the corresponding projections of $M^x, M^y$ (this is because two of the spins in each case have zero component on the given axes $x,y$). Below, we focus on a `cut' through the ${\bf G} = 002$ zone centre such that ${\bf q} = hh0$ in reduced units and the observed differential cross sections are 
\begin{align}
\frac{\textup{d}\sigma^{\rm NSF}_{hh2}}{\textup{d}\Omega} &= S^{xx}_{hh0},\notag\\
\frac{\textup{d}\sigma^{\rm SF}_{hh2}}{\textup{d}\Omega} &= S^{yy}_{hh0} \cos^2\theta + S^{zz}_{hh0}  \sin^2\theta,\label{sf}
\end{align}
where $\theta$ is the angle between ${\bf Q} = {\bf G}+ {\bf q}$ and 
${\bf G} = 002$. Fortunately, the phase factors of the $zz$ component act to suppress the second term in~\cref{sf} such that it may be neglected up to $h \approx 0.5$. The net result is that the SF and NSF channels measure the longitudinal (L) and transverse (T) eigenvalues of the magnetization structure factor to an excellent approximation. We can therefore conveniently compare $\textup{d} \sigma^{\rm NSF}_{hh2}/\textup{d}\Omega$  with $S^{\rm T}_{hh0}$ and 
$\textup{d} \sigma^{\rm SF}_{hh2}/\textup{d}\Omega$  with $S^{\rm L}_{hh0} \cos^2\theta = S^{\rm L}_{hh0} \times 4/(4+2h^2)$ in this range.

 \medskip

\noindent
{\bf Near Neighbor Spin Ice --} The scattering function of near neighbor spin ice (NNSI) may be calculated from the free energy functional~\cite{Bramwell_PhilTrans}: 
\begin{equation}\label{GNNSI}
G[{\bf M}] = \mu_0 \int\left(  \frac{{\bf M}^2}{2\chi} + \frac{\xi^2 (\nabla\cdot{\bf M})^2}{2 \chi}  - {\bf H}\cdot{\bf M}\right) \textup{d}^3r,
\end{equation}
which comes from an expansion of the free energy in powers of magnetization ${\bf M}$. Its first term in the susceptibility $\chi$ and its third term in the field ${\bf H}$ are common to all classical spin systems, while the second term in $(\nabla\cdot{\bf M})^2$ expresses the free energy cost of divergence and is the lowest order gradient term for ice systems~\cite{YA,YAM}.  The Euler-Lagrange equation 
$
{\bf M} - \xi^2 \nabla\left(\nabla \cdot{\bf M}\right) = \chi {\bf H},
$
may be solved in Fourier components to give the wave vector dependent susceptibility $\chi({\bf q})$ and, by the classical fluctuation dissipation theorem, the eigenvalues of the scattering tensor: 
\begin{equation}\label{sq}
S^{\rm T}(q) = \chi T/3C~~~~~~~~~ S^{\rm L}(q) = \frac{\chi T/3C}{1 + \xi^2 q^2}. 
\end{equation}
Here T/L indicate transverse/longitudinal to the wavevector ${\bf q}$. These depend on the susceptibility $\chi$ and `diffusion' length $\xi$~\cite{RRB}. 

The susceptibility may be expressed as $\chi = \gamma(T) C/T$, with $C = \mu_0 m^2 /(3 k_{\textup{B}} v_0/2)$, where $m$ is the rare earth moment and $v_0=v/2$ is the volume per tetrahedron.  The moment is related to the monopole charge $Q$ by $Q = 2 m/a$ where $a$ is the diamond lattice constant~\cite{CMS}. The factor $\gamma(T)$ has been calculated by Jaubert {\it et al.}~\cite{Jaubert} in a Husimi tree approximation:
\begin{equation}\label{Jaubert}
\gamma(T) = \chi T/C = \frac{2 (1 + e^{2 J_{\rm eff}/T})}{2 + e^{2 J_{\rm eff}/T} + e^{-6 J_{\rm eff}/T}}, 
\end{equation}
where $J_{\rm eff}$ is the effective exchange defined above~\cite{Bramwell_Gingras}. 

The diffusion length $\xi(T)$ depends on the densities of single and double charge monopoles (which have no Coulomb interaction in NNSI). We make a low temperature approximation by setting the chemical potential of the double charge monopoles infinite. We write the entropy~\cite{Ryzhkin, Kaiser} in terms of the densities of positive and negative monopoles, pick out a component of the free energy $G = -\mu{n_+} - \mu{n_-}  - T S$ (with $\mu = 2 J_{\rm eff}$) and then substitute
$n_+ \rightarrow (n + \delta_n)/2$ and $n_- \rightarrow (n - \delta_n)/2$ where $n$ is the equilibrium monopole density and $\delta_n$ a charge fluctuation. A Taylor expansion of the free energy in powers of $\delta_n$, followed by setting the linear term to zero gives the equilibrium density~\cite{Ryzhkin}:
\begin{equation}\label{n}
n = \frac{4 e^{-2 J_{\rm eff}/T}}{3 + 4 e^{-2 J_{\rm eff}/T}},
\end{equation}
and a second order fluctuation term $G'' =  \mu_0 kT (\delta_n)^2/2 n v_0$.  The latter may be related to~\cref{GNNSI} using
$
(\delta_n)^2 =(\nabla \cdot {\bf M} )^2/(Q/v_0)^2
$
to give~\cite{Bramwell_PhilTrans}:
\begin{equation}\label{xi}
\xi^2 = \frac{\chi kT v_0 }{\mu_0 Q^2 n}.
\end{equation}
\cref{sq,Jaubert,n,xi} combine to give an analytic calculation of the full scattering function tensor. This is a low temperature approximation: for ${\rm Ho_2Ti_2O_7}$ parameters ($2 J_{\rm eff} = -\mu =  3.8$ K), it should be valid at $T =2 $ K and below, where double charge monopoles are very scarce. 

\begin{figure}[!htb]
		\centering
  		\resizebox{\hsize}{!}{\includegraphics{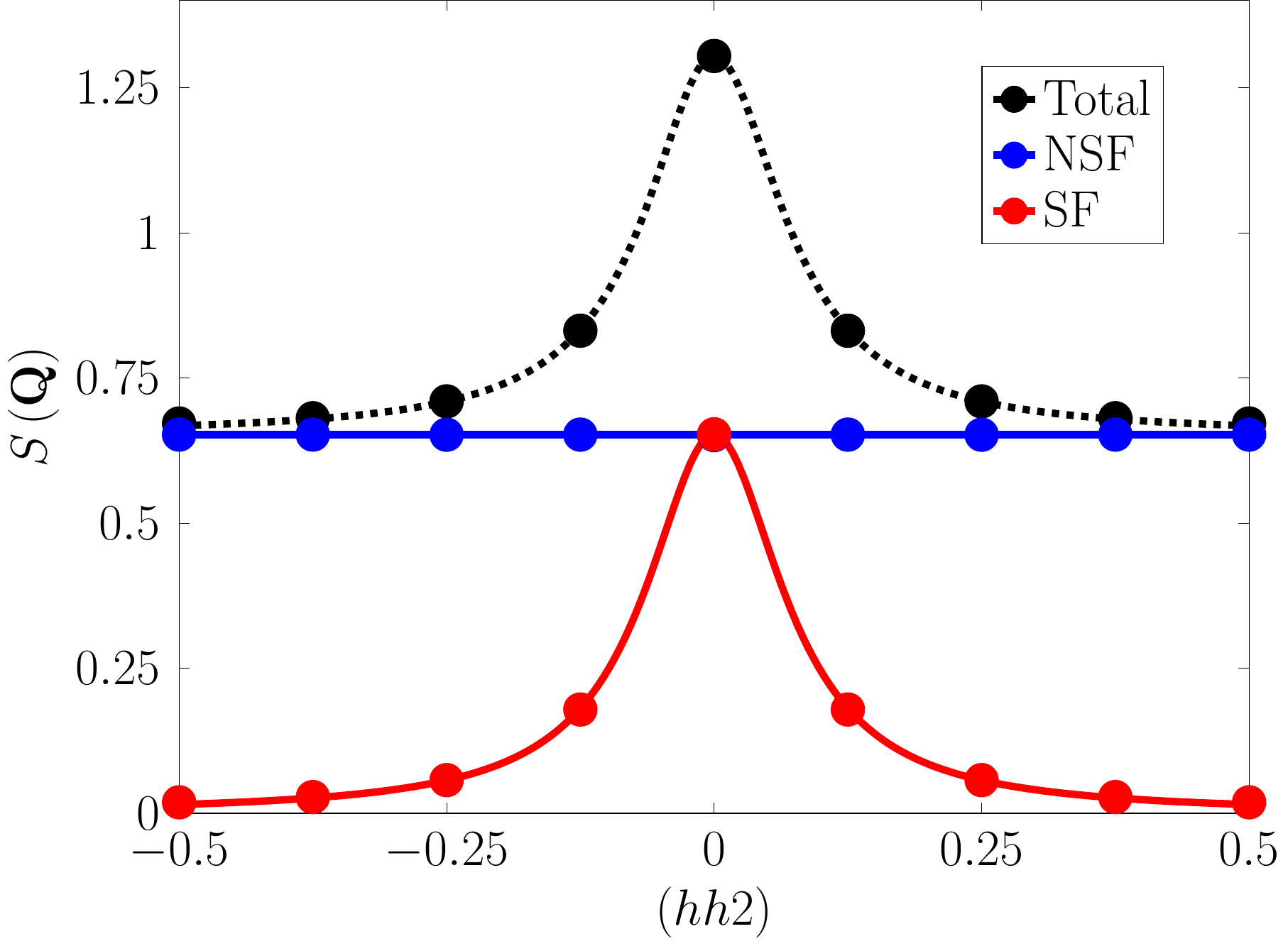}}
 		 \caption{Test of the accuracy of our analytic theory of nearest neighbor spin ice (NNSI, periodic boundary conditions). The pinch point profile at $T = 1$ K with ${\rm Ho_2Ti_2O_7}$ parameters. Numerical simulation (points) versus the analytical theory developed in this paper (lines). The comparison of analytical theory and simulation is observed to be essentially perfect.}
 		 \label{NNSIfig}
\end{figure}

In~\cref{NNSIfig} we compare the calculated scattering functions of NNSI with the numerically simulated one: the parameter-free comparison is excellent, showing that the continuum theory is a nearly exact description of NNSI over the wave vectors considered. This gives us great confidence that the theory can be adapted to describe dipolar spin ice (DSI) as well. 

\medskip
\noindent
{\bf Inclusion of dipolar interactions --}
Given the excellent comparison of theory and simulation for NNSI, we seek to describe DSI by a natural extension of the NNSI theory. In particular, we simply add a term to~\cref{GNNSI} to describe the magnetostatic energy:
\begin{equation}
U_{\rm mag} = -\frac{\mu_0}{2}\int {\bf M}\cdot\tilde {\bf H}~\textup{d}^3r, 
\end{equation}
where $\tilde{\bf H}({\bf r})$ represents the local internal far field arising from the dipoles at points ${\bf r}'$ in the sample. Treating $\tilde{\bf H}$ as an independent variable, the Euler-Lagrange equation becomes:  
\begin{equation}
{\bf M} - \xi^2 \nabla\left(\nabla \cdot{\bf M}\right) = \chi({\bf H} + \tilde{\bf H}),
\end{equation}
which can be solved for the structure factor as above if $\tilde{\bf H}$ can be expressed as a function of ${\bf M}$. 

It is well known (see, for example, Ref.~\cite{Morrish}) that the field at a point external to a system of dipoles (e.g. at ${\bf r}$ where the dipoles are at ${\bf r}'$) may be expressed as a sum over the fields arising from surface magnetic charge density $\sigma = {\bf M}\cdot\hat{\bf n}$ (where $\hat{\bf n}$ is the unit vector normal to the surface) and volume magnetic charge density $\rho = -\nabla\cdot{\bf M}$. Without yet being explicit about surface effects, we write 
\begin{equation}
\tilde{\bf H}({\bf r}) = -\frac{\mu_0}{4 \pi}\nabla \int \frac{\rho({\bf r}')}{|{\bf r-r}'|} \textup{d}^3 r'.
\end{equation}
By Helmholtz' theorem, the dipolar field is then equal to the irrotational (longitudinal) component of the magnetization, $\tilde{\bf H} = - {\bf M}^{\rm L}$. Writing ${\bf M} = {\bf M}^{\rm L} + {\bf M}^{\rm T}$ thus renders the 
dipolar correction entirely local and we may proceed as above to find the structure factors:
\begin{equation}\label{long1}
S^{\rm T} =\gamma/3~~~~~~S^{\rm L} =  \frac{\gamma/3}{(1+\chi)+q^2 \xi^2}.
\end{equation}
These are the same as~\cref{sq} but with the the longitudinal structure factor suppressed by a factor $1/(1+ \chi)$ at the zone centre. Also, the characteristic length becomes the Debye length $\xi/\sqrt{1+ \chi}$ rather than the diffusion length $\xi$~\cite{RRB}. In addition (not shown in~\cref{long1}) there is a delta function at the zone centre: this is elucidated for the case of different boundary conditions below. 

It has been shown~\cite{Bramwell_NComms} that treating the dipolar field in this way is equivalent to Onsager's cavity construction~\cite{Onsager} for dipolar fields, where the magnetic charge resides on the surface of a spherical cavity cut in a continuous polarizable medium. Alternatively, we can see that the dipolar integral is equivalent to a discrete summation over the effective magnetic monopoles of spin ice. These two different physical pictures give the same result only if the fields are unscreened~\cite{Bramwell_NComms}. Hence we refer to this as the `unscreened' model going forwards. 

\begin{figure}[!htb]
		\centering
  		\resizebox{\hsize}{!}{\includegraphics{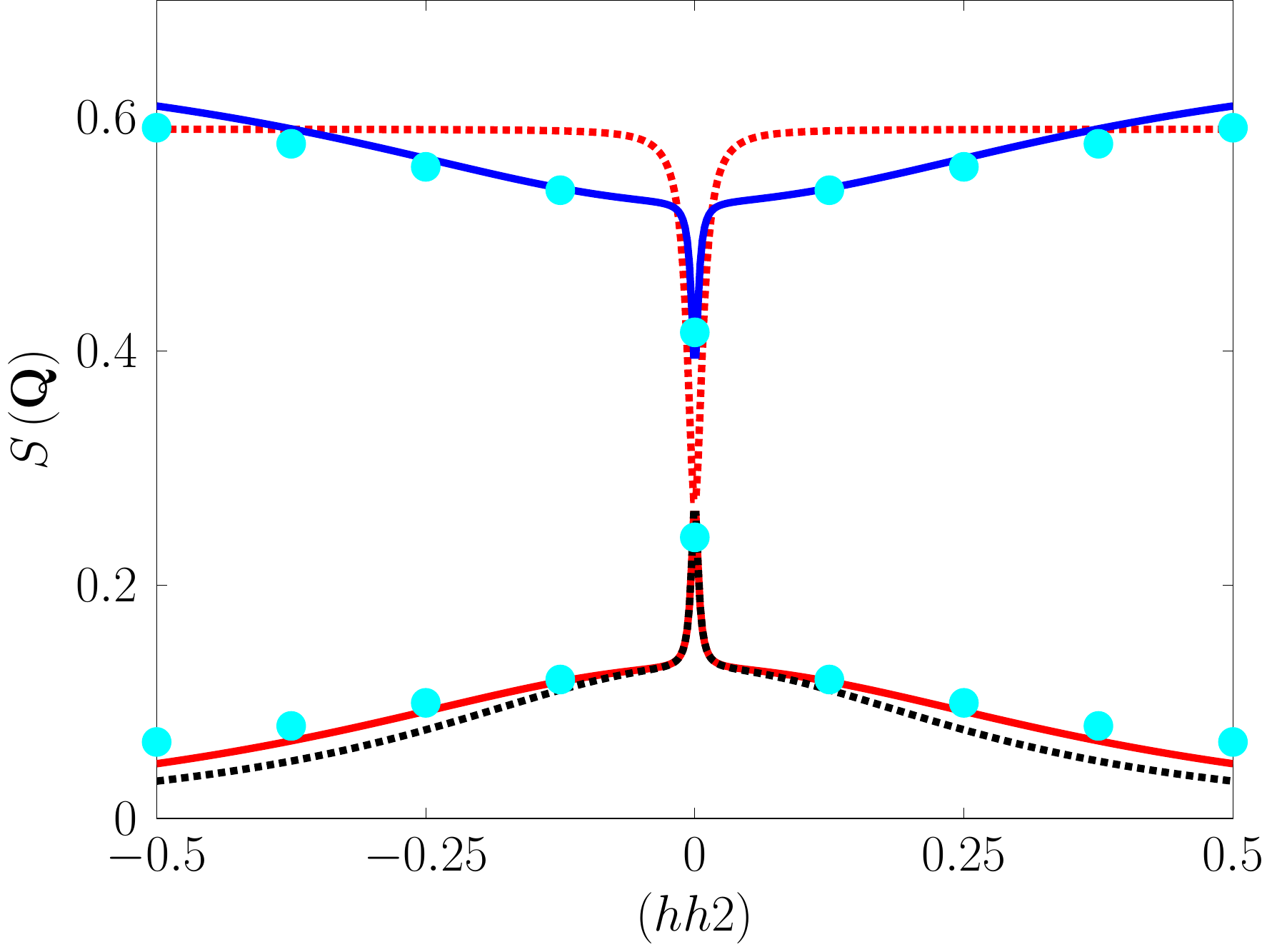}}
 		 \caption{Dipolar spin ice (DSI) at $T = 2$ K with ${\rm Ho_2Ti_2O_7}$ parameters and spherical boundary conditions. Comparison of numerical simulation (points) and the {\it unscreened} analytical theory presented in this paper (lines). Red lines: unmodified analytical theory (\cref{long1}). Blue line: NSF scattering generated with $S^{\rm T} \sim 2/3 -S^{\rm L}$ ) (see text). Dotted line: result using monopole density replacing the `bare' defect density (see text). The figure shows that the simplest possible extension of the near neighbor spin ice theory to include dipolar interactions captures the numerical data to a close approximation.}
 		 \label{DSIfig}
\end{figure}

Following~\cref{sf} and the discussion thereafter, we may once again compare $\textup{d} \sigma^{\rm NSF}_{hh2}/\textup{d}\Omega$  with $S^{\rm T}_{hh0}$ and 
$S^{\rm SF}_{hh2}$  with $S^{\rm L}_{hh0} \cos^2\theta$ up to $h=0.5$. \cref{DSIfig} compares the DSI simulation with the unscreened dipolar calculation at $T=2$ K. We see that the simulated SF scattering function is immediately well described by the unscreened model, the NSF slightly less well.  Since $\chi$ is large (e.g. $\chi \approx 6$ at $T = 2$ K~\cite{Bovo_Curie}), the Lorentzian peak becomes very broad and flat compared to that of NNSI.

There are two small modifications to the unscreened model that are worth considering. First, we notice that the NSF scattering is better described by $S^{\rm T} \sim 2/3 -S^{\rm L}$ (see~\cref{DSIfig}). As discussed in Ref.~\cite{Bramwell_NComms}, this correction is both plausible, because it satisfies the total moment sum rule, and justifiable, because the longitudinal fields were altered independently of the transverse. Second, the scattering function~\cref{long} apparently depends on $2J_{\rm eff}$ rather than the magnetic monopole chemical potential, which exceeds $2J_{\rm eff}$ in magnitude owing to the Coulomb interaction between monopoles.  This difference increases the equilibrium defect density $n$,  altering the specific heat~\cite{Kaiser}. As~\cref{long} depends on $n$ through $\xi$ (\cref{n,xi,long}), it might be appropriate to replace $n$ by the true equilibrium density~\cite{Kaiser}. A fit to the ensuing function is tested in~\cref{DSIfig} but in fact it makes the description of the DSI simulation slightly poorer.

\medskip
\noindent
{\bf Boundary conditions and screening --}
To account for different boundary conditions, we add 
a `self field'~\cite{Griffiths} to the dipolar field which generates the correct average when the field is summed over a ellipsoidal volume: in this way the demagnetizing factor $N_{\rm d}$ ($= 1/3, 0$ for spherical and Ewald boundaries respectively) naturally enters the equations, obviating the need to treat surface charge explicitly. This factor is conveniently introduced by treating the dipolar integral as a sum over thermally generated monopoles and allowing these to screen each other as in a Debye-H\"uckel gas~\cite{Kaiser}. In this approximation, the dipolar field is not simply equal to minus ${\bf M}^{\rm L}$, but in addition has a longer-range harmonic component. At the very least, such a component is expected from the demagnetizing field that arises from the surface charge (surface monopoles). It was suggested in Ref.~\cite{Bramwell_NComms} that it could also arise from thermally generated monopoles at mesoscopic distances~\cite{Bramwell_NComms}, but we will examine the validity of this proposal subsequently. 

We first consider the Debye-H\"uckel equation for the potential $\phi$:
\begin{equation}
\nabla \cdot \nabla \phi({\bf r}) + \kappa^2 \phi({\bf r})= 0,
\end{equation}
which applies outside a region of size $\sim a$ at the origin. By taking the gradient of this equation and introducing a delta function to extend it to the origin we find for the field at all ${\bf r}$: 
\begin{equation}
- \nabla\left(\nabla \cdot\tilde{\bf H}({\bf r})\right) - \kappa^2 \tilde{\bf H}({\bf r}) = \left[\nabla\left(\nabla\cdot{\bf M}\right) + \kappa^2 N_{\rm d}{\bf M}({\bf r}) \right]\delta ({\bf r}),
\end{equation}
where $-N_{\rm d}{\bf M}({\bf r})\delta({\bf r})$ is the self-field. 

The Debye-H\"uckel solution $\kappa\rightarrow \kappa_{\rm DH} $ then gives the Greens function solution for the more general equation where charge density is widely spread: 
\begin{align}
&- \nabla\left(\nabla \cdot\tilde{\bf H}({\bf r})\right) - \kappa^2 \tilde{\bf H}({\bf r}) =\notag \\
&\int\left[ \nabla\left(\nabla\cdot{\bf M}({\bf r}')\right) + \kappa^2N_{\rm} d{\bf M}({\bf r}') \right] \,\delta({\bf r-r}')\, \textup{d}^3 r'.   
\end{align}

In Fourier components the field becomes
\begin{equation}
\tilde{\bf H} ({\bf q}) =-\left({\bf q}\left({\bf q}\cdot {\bf M} ({\bf q})\right)+ \kappa^2 N_{\rm d}{\bf M} ({\bf q})\right)/(\kappa^2+q^2),
\end{equation}
and the solution of the Euler-Lagrange equation is then:
\begin{equation}\label{long}
S^{\rm L} = \frac{(\gamma/3) \left(\kappa^2+q^2\right)}{\kappa^2(\chi N_{\rm d}+1) +q^2(1+\chi +\kappa^2 \xi^2)+q^4 \xi^2}
\end{equation}
\begin{equation}
S^{\rm T} = \frac{(\gamma/3) \left(\kappa^2+q^2\right)}{\chi \kappa^2 N_{\rm d}+(\kappa^2+q^2)}. 
\end{equation}
Setting $\kappa \rightarrow 0$ in~\cref{long} results in the unscreened scattering functions (\cref{long1}). In~\cref{DSIfig}, to compute the scattering functions, we left $\kappa$ small but finite, which broadens the delta function slightly to render it visible to the eye. 

\begin{figure}[!htb]
		\centering
  		\resizebox{\hsize}{!}{\includegraphics{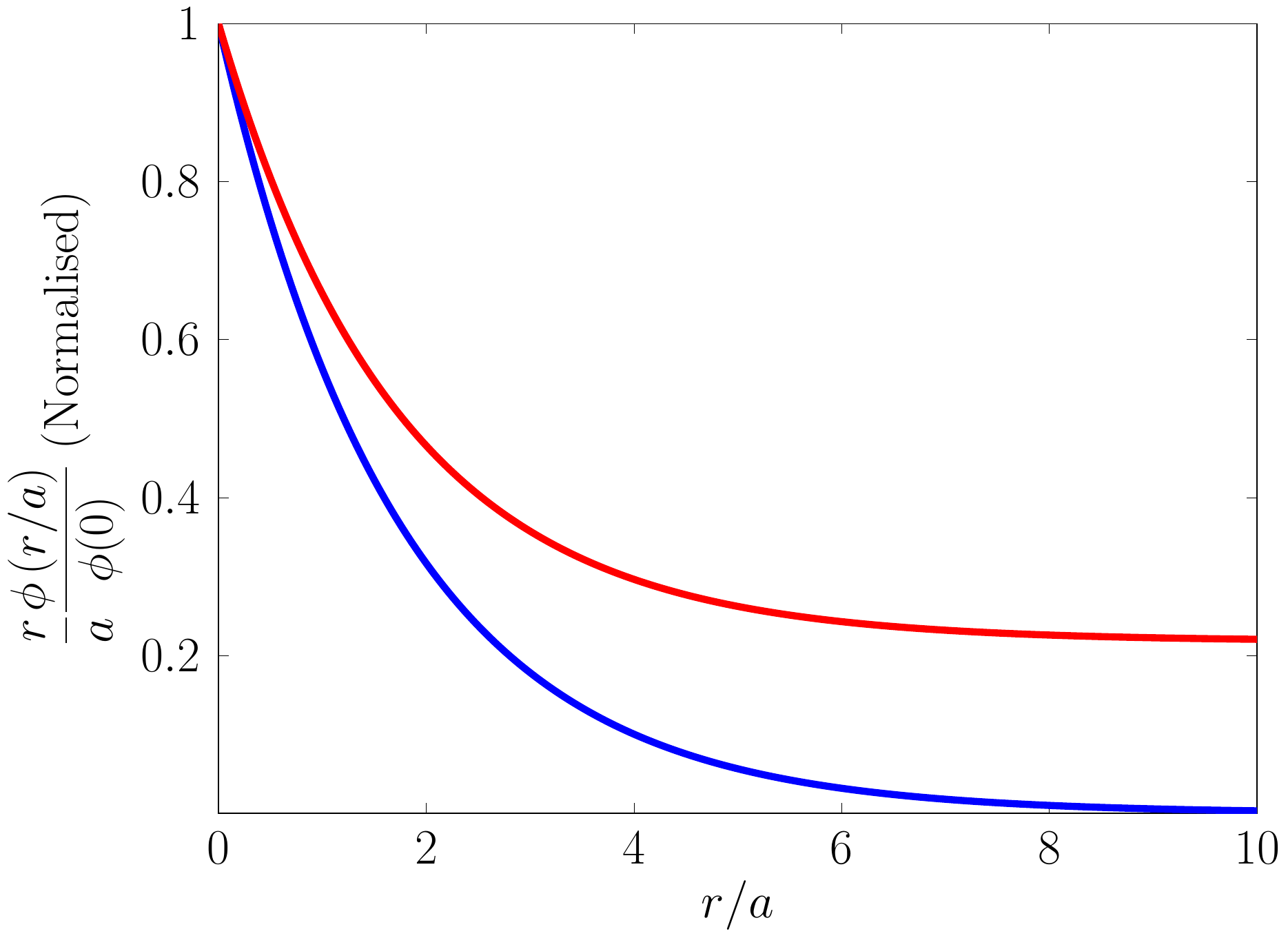}}
 		 \caption{Magnetic monopole potential calculated using ${\rm Ho_2Ti_2O_7}$ parameters at $T=2$ K. The figure shows $r \phi(r)$ normalized to its value at $r=0$ (here $a$ is the lattice constant of the diamond lattice inhabited by the monopoles~\cite{CMS}). Blue line: the standard Debye-H\"uckel potential $\phi(r)\sim e^{-r}/r$ which is screened to zero at long distance.  Red line: the monopole potential in spin ice~\cite{RRB}, revealing a power-law tail $\phi(r) \sim 1/r$ and hence an unscreened contribution at long distance. The unscreened contribution is the cause of the infinitely sharp pinch points in dipolar spin ice (DSI). It is confirmed in simulation (\cref{DSIfig}) but is not observed in experiment (\cref{PPP}).}
 		 \label{potentialfig}
\end{figure}

It might seem surprising that we have to remove the screening ($\kappa \rightarrow 0$) in order to describe the numerical data: after all, it is well established from analysis of the specific heat, that monopoles are screened with a finite $\kappa$~\cite{Kaiser}. In fact, it was shown in Ref.~\cite{RRB} that the monopole potential $\phi$ consists of both a screened part and an unscreened part (see~\cref{potentialfig}).  The energy and hence specific heat is determined by the short range, screened, part of the potential whereas the pinch point is determined by the long ranged, unscreened part. Our analysis using the Poisson-Boltzmann equation above forces the potential into a perfectly exponential decay. Hence Debye-H\"uckel theory can only describe the pinch point if the screening length $\kappa^{-1}$ is set to be infinite, recovering the Poisson equation. In contrast, to calculate the specific heat~\cite{Kaiser}, it incurs negligible error to neglect the power law tail of the potential, and equate $\kappa$ with the Debye length of the monopole fluid.

\begin{figure}[!htb]
		\centering
  		\resizebox{\hsize}{!}{\includegraphics{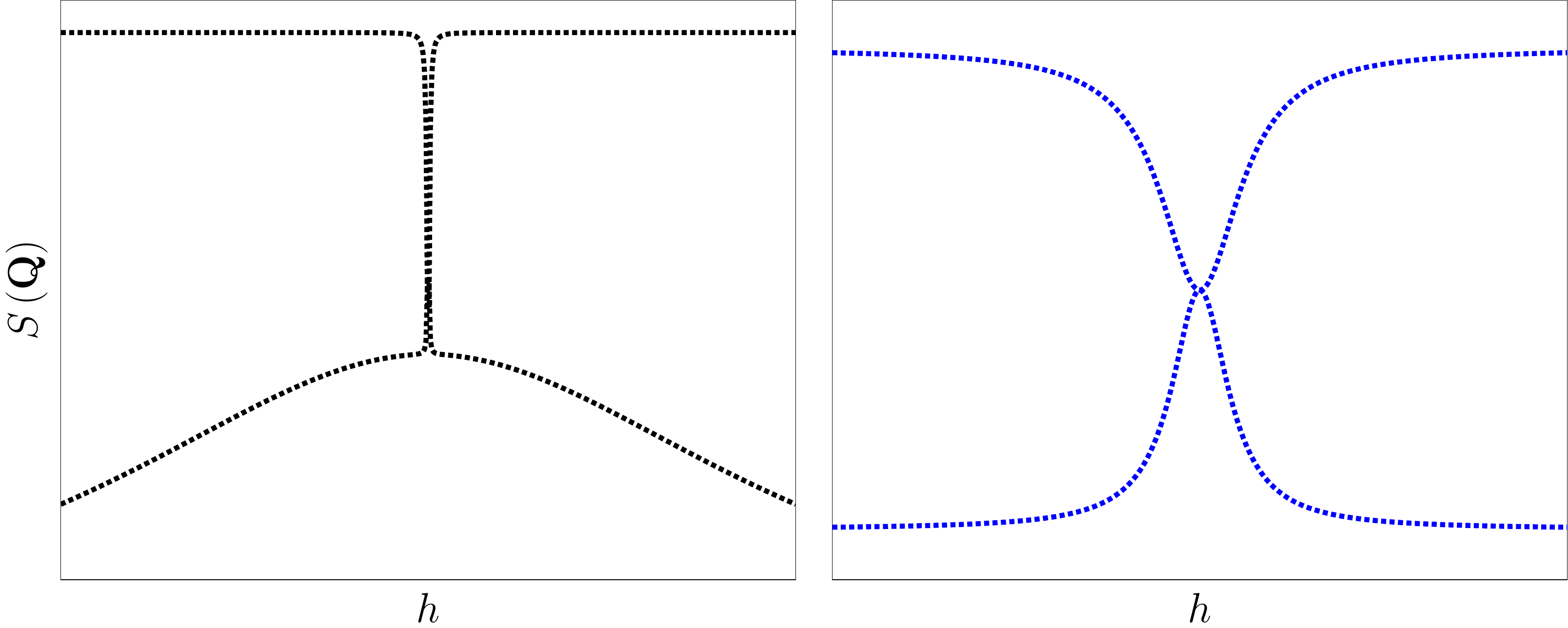}}
 		 \caption{Limiting line shapes in our more general analytical theory that allows for screening of dipolar fields by magnetic charge density. Left: the case of infinite screening length ($1/\kappa = \infty$) and finite defect diffusion length $\xi$. Right: the case of finite screening length ($1/\kappa$) and zero diffusion length $\xi=0$. It is seen that the screening length $1/\kappa$ controls the sharpness of the central peak, while the defect diffusion length controls the flatness of the `background'. The completely flat background in the right hand feature corresponds to the `harmonic phase' of Ref.~\cite{Bramwell_NComms}. }
 		 \label{lineshapefig}
\end{figure}

\cref{long} suggests further possibilities to test. Varying $\kappa$ and $\xi$ generates a family of curves, as illustrated in~\cref{lineshapefig}. We see that finite $\kappa$ broadens the central delta function into a Lorentzian in the longitudinal channel, plus a inverse Lorentzian `dip' in the transverse channel. We consider the model in which $\kappa$ retains its identity as the inverse Debye length for single and double-charge magnetic monopoles. This may be calculated as in Ref.~\cite{Kaiser}, which has been shown to give a very accurate calculation of specific heat in both simulation and experiment. Additionally, setting $\xi = 0$ results in an entirely flat background to this central Lorentzian. This flat background is the scattering function of a dipolar fluid in the Onsager picture - the `harmonic phase' of Ref.~\cite{Bramwell_NComms}.  Hence the right hand curve of~\cref{lineshapefig} is the scattering function of dipoles screened by magnetic monopoles, the `screened dipolar fluid'~\cite{Bramwell_NComms}. 

\begin{figure}[!htb]
		\centering
  		\resizebox{\hsize}{!}{\includegraphics{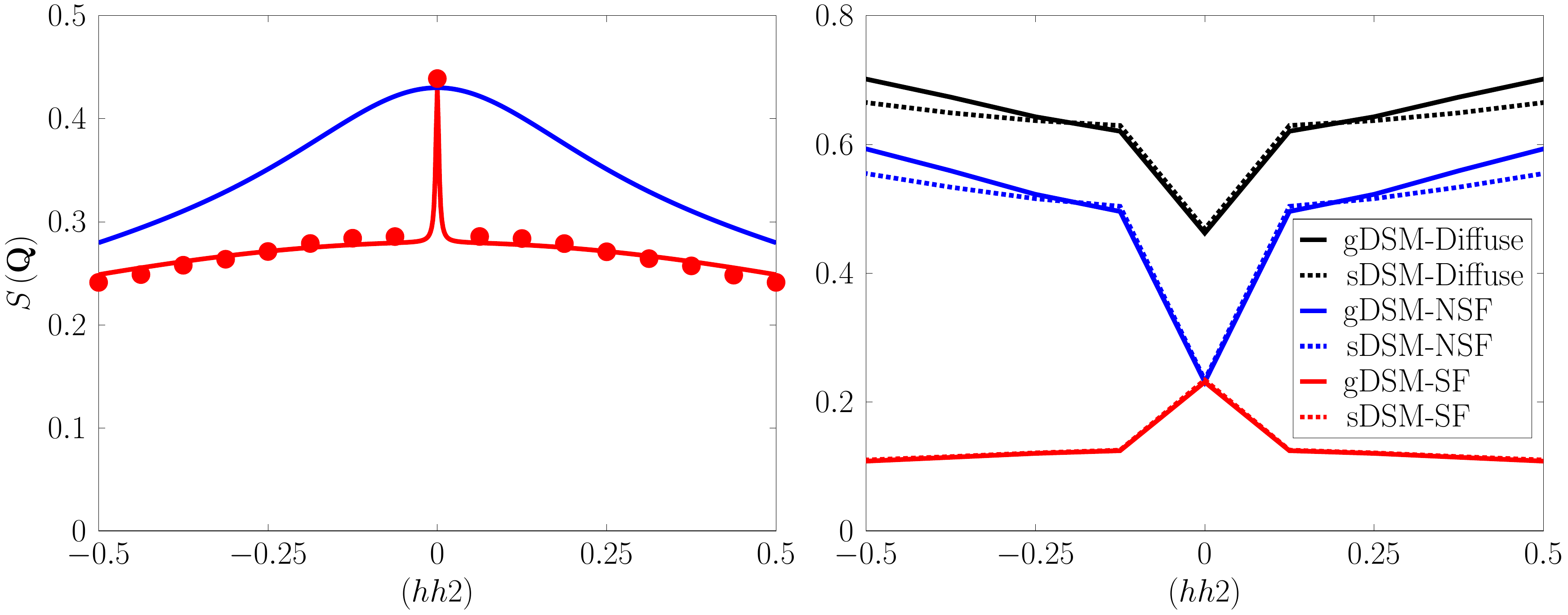}}
 		 \caption{Left. A high resolution simulation of DSI at $T=10$ K to test for the sharpness of the central peak component. Points: simulated data. Lower lines: the unscreened line shape. Upper line: the corresponding NNSI line shape for comparison. There is no broadening of the central peak within the resolution of the simulation. Right. Comparison of simulations of the dipolar spin ice model with single exchange constant ($s$DSM) with the generalized dipolar model ($g$DSM, ${\rm Dy_2Ti_2O_7}$ parameters~\cite{Yavorskii}). Simulated points spaced by 0.125 are linked by full lines as guides to the eye. Differences between $s-$ and $g$DSM are observed to be confined to the wings of the scan, showing that the detailed tuning of short-range exchange constants does not significantly affect the pinch point profile near the zone center.}
 		 \label{highresfig}
\end{figure}

\medskip
\noindent
{\bf Infinitely sharp pinch points in DSI simulation--}
We have already established (\cref{DSIfig}) that the unscreened dipolar model describes our simulation well, but can we actually rule out the screened model (finite $\kappa$)? To test this, we examine the simulation at a much higher temperature (10 K) where $\kappa$ should be sufficiently large to make the central Lorentzian (broadened delta function) easily visible within the resolution of our simulation. The result, shown in~\cref{highresfig}, does indeed rule out any pinch point broadening on the expected scale. (Note that, to obtain the fit to the unscreened model, we adjusted $\xi$ slightly to account for double charge monopoles).  We also note that our simulation largely rules out the screened dipolar fluid model of Ref.~\cite{Bramwell_NComms}. The `harmonic phase' (flat background), is, however, present to a certain approximation, given that the `diffusion Lorentzian' is very broad and flat at higher temperatures. 

The physical origin of the infinitely sharp pinch points in dipolar spin ice simulations may now be traced to the
unscreened part of the monopole potential calculated in Ref.~\cite{RRB} and already mentioned. This feature was also noted as a dipolar tail in the correlation function in Ref.~\cite{Sen}.  Referring to~\cref{potentialfig}, the long-ranged unscreened part of the potential gives rise to a `gap' between longitudinal and transverse eigenvalues of the scattering tensor at $q\rightarrow 0^+$, and hence the zone centre delta function that we observe. However, in view of the experimental result (see below), it was suggested in Ref.~\cite{RRB} that the gap is unrealistic. This is an issue that we consider further below.  

\section{The pinch point paradox}\label{PPP}
In the preceding Section we have established a parameterized form of the pinch point profile that accurately and quantitatively describes high resolution simulations of the dipolar spin model. This is valuable for making a bridge to experiment as through parameter adjustment it allows for other physical possibilities and physical insight. As already advertised, it is in this confrontation of theory and experiment that the pinch point paradox arises.  

The pinch point lineshape of polarized neutron scattering in ${\rm Ho_2Ti_2O_7}$ spin ice has been independently studied by Fennell {\it et al.}~\cite{Fennell} and by Chang {\it et al.}~\cite{Chang}. Both sets of authors found that it can be described by a central Lorentzian plus a temperature-dependent flat (i.e. $q$-independent) component at all temperatures. An explanation of the flat component was put forward by one of us in Ref.~\cite{Bramwell_NComms} in terms of the screened dipolar fluid model described above, i.e. $\xi = 0$, finite $\kappa$ as calculated in Ref.~\cite{Kaiser}. This quantitatively produced the observed temperature dependence of the flat component as well as producing a Lorentzian central peak (note that in subsequent figures, the `flat' background is not completely flat because of the $\cos^2\theta$ factor discussed above). However, the screened model of Ref.~\cite{Bramwell_NComms} is already ruled out as a model for the DSI simulations. 

\begin{figure}[!htb]
		\centering
  		\resizebox{\hsize}{!}{\includegraphics{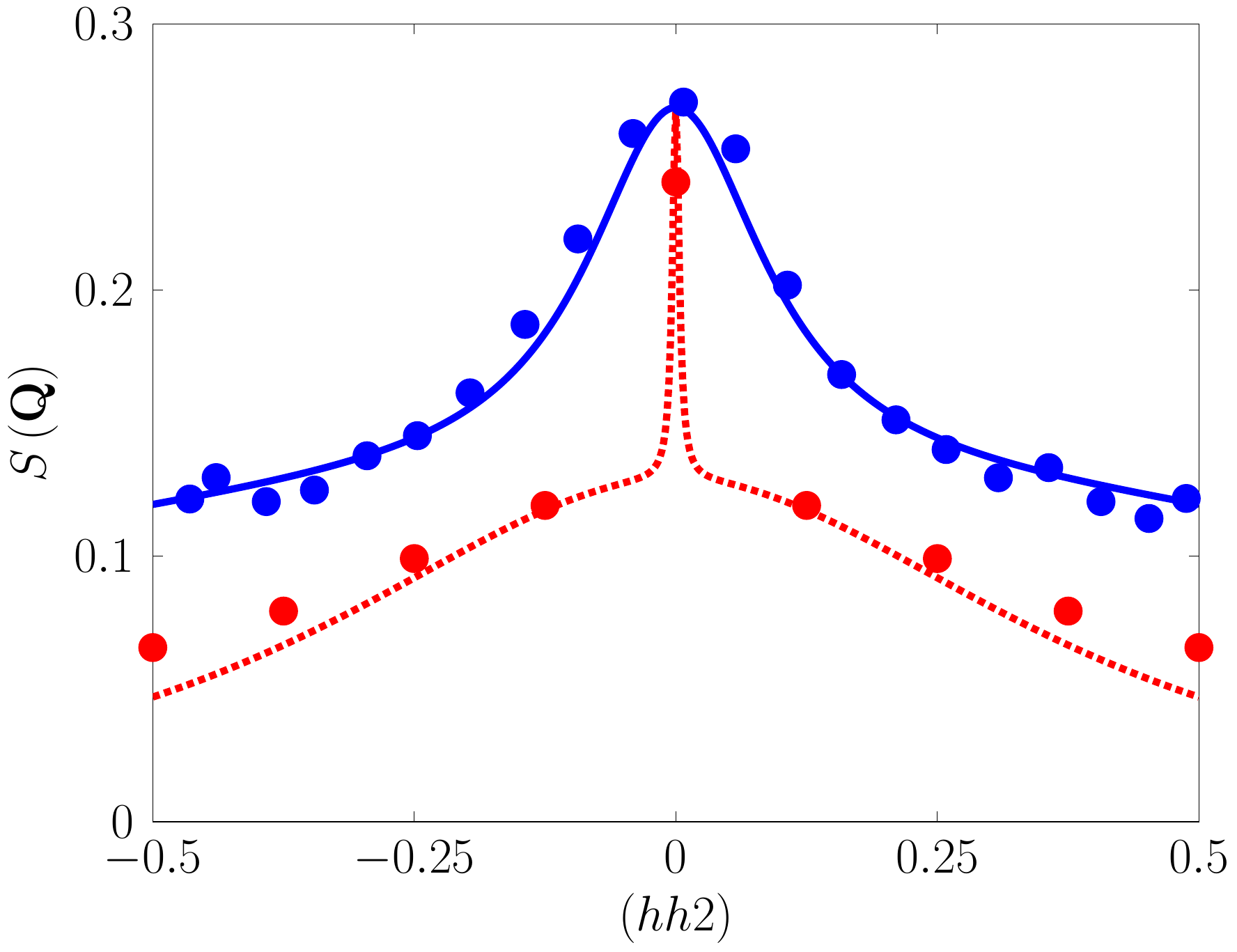}}
 		 \caption{The pinch point paradox in spin ice. Comparison of experimental data and numerical data with the limiting forms of the analytical theory. Chang {\it et al.}'s SF data~\cite{Chang} (blue points) and our simulation (red points, spherical boundary conditions) at $T = 2$ K. The experimental data have been scaled to fit the theory (lines) at $002$. Blue line: screened theory and red line: unscreened theory. The fact that experiment and simulation are only consistent with {\it different} limiting cases of the {\it same} analytical theory (\cref{lineshapefig}) is the essence of the pinch point paradox.}
 		 \label{paradoxfiga}
\end{figure}

The essence of the pinch point paradox is shown in~\cref{paradoxfiga} where we compare the SF scattering at $T = 2$ K measured by Chang {\it et al.}~\cite{Chang} with the DSI simulation, and the screened and unscreened theories (the latter for $\xi = 0$, the dipolar fluid case~\cite{Bramwell_NComms}). The experimental data were scaled such that the $q = 0$ point is coincident with the theory: i.e. we assume that theory and experiment are consistent as regards the bulk susceptibility (this is an approximation, but a reasonably accurate one). We see very clearly that the experiment rules out the unscreened DSI but is remarkably well described by the screened dipolar fluid. Indeed, the screened line shape can be transformed~\cite{Bramwell_NComms} into independent flat and Lorentzian components, where the Lorentzian component has the same width parameter as NNSI. This is in perfect agreement with the analysis of Chang {\it et al.} who fitted this data using Monte Carlo simulations of NNSI plus an arbitrary flat background: our theory eliminates the adjustable parameter used by these authors. They also drew a central Lorentzian through the points of their simulations of DSI -- a reasonable extrapolation in view of knowledge available at the time, but one that we have now shown to be incorrect through higher resolution simulations. Because DSI has a delta function rather than a Lorentzian central peak, the discrepancy between simulation and experiment for spin ice is a very large one -- yet both simulation and experiment should {\it a-priori} be considered correct, hence the paradox. 

\begin{figure}[!htb]
		\centering
  		\resizebox{\hsize}{!}{\includegraphics{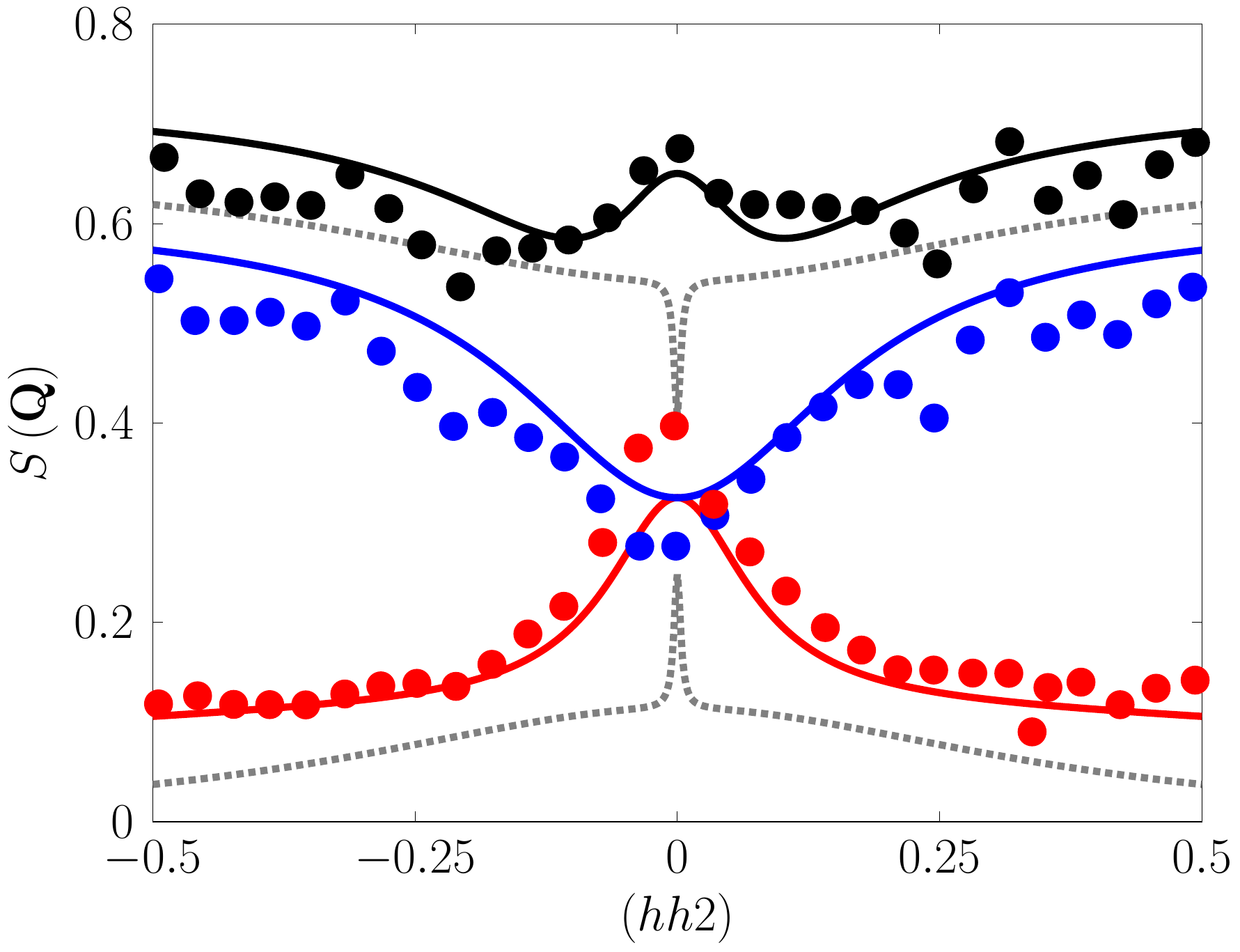}}
 		 \caption{Further comparison of experimental data and numerical data with the limiting forms of the analytical theories at $T=1.7$ K. A comparison for the SF (lower, red),
 		 NSF (middle, blue) and total (upper, black) data of Fennell {\it et al.}~\cite{Fennell}. The red blue and black lines are the screened theory. The experimental data have been scaled such that the total scattering approximately fits the full black line. The dotted black lines are the unscreened theory that has been shown to described DSI extremely well (\cref{DSIfig}). It is concluded that numerics and experiments are again described by different limiting cases of the same theory (see~\cref{lineshapefig}). Note that the demagnetizing factor in the screened theory has been adjusted to better fit the data in the centre of the scan (see text).}
 		 \label{paradoxfigb}
\end{figure}

In the case of the 1.7 K SF, NSF and total neutron scattering data of Fennell {\it et al.}~\cite{Fennell}, the DSI again fails badly, while the screened dipolar fluid is again close to the experimental data (see~\cref{paradoxfigb} -- the fit has been improved slightly by adjusting the demagnetizing factor to ~0.2). Notably, the screened dipolar fluid model captures the striking `zone center' dip  in the SF scattering as well as the small `zone center bump' in the total scattering. In general, we found that the screened model outperforms the unscreened one at all temperatures, in every case describing the flat (harmonic phase) background well, though at higher temperatures overestimating the Lorentzian width by a significant factor (e.g. $\sim 5$). Despite this, we can certainly conclude that, in the temperature range where we are most sure of the theory, the screened dipolar fluid model is fully consistent with experiment, while the DSI simulation and unscreened models are ruled out as descriptions of experiment. 

\section{Discussion and Conclusion}

Our study exposes two significant results: (i) the pinch point profile in the 
dipolar spin ice simulation is infinitely sharp, and (ii) experiment disagrees with simulation in that the profiles are broad, rather than sharp. Yet both are described quantitatively by different limits of the same theory.

To elaborate on (i), we recall that the subtle long-ranged correlations in spin ice are entirely revealed by neutron polarization analysis of the pinch point profile, as explained in the introduction (section I). In the near neighbor spin ice model the pinch points are broad, but addition of the long range dipole-dipole interaction causes them to be infinitely sharp with a delta function singularity at $q = 0$~\cite{Sen,RRB}. The structure factor (inverse) tensor remains an oblate spheroid (see~\cref{figure:1}\,c,d), except precisely at $q = 0$, where it becomes a sphere, in keeping with the cubic space symmetry~\cite{Nye}.

As dipolar spin ice represents a model Coulomb gas, this raises an interesting and topical point. The Debye length for magnetic monopole interactions $( = \xi/\sqrt{1+\chi})$ is always finite, and any long-ranged correction to the exponential screening is essentially negligible as regards monopole correlations or specific heat. However the field-screening length ($= 1/\kappa$) is, at the same time, {\it divergent}. Thus dipolar spin ice -- a polarizable Coulomb gas that is relevant even as a model of water ice~\cite{Ryzhkin_DSI} -- presents a very different picture as regards charge correlations and field correlations: the screening length for fields diverges while the Debye length remains finite. This is not a contradiction as it depends on how a particular experimental measure picks out the details of what is a rather subtle and long-ranged correlation function~\cite{RRB} (see~\cref{potentialfig}). In this context, we may make a connection with an interesting recent discussion of the experimental observations of `underscreening' in dense ionic liquids~\cite{Lee2015,Lee}. We see that dipolar spin ice affords a tangible example where the screening length diverges, while the Debye length remain constant. This suggests that the observations of Ref.~\cite{Lee2015,Lee} may be consistent with the dense ionic liquid having a stable polarization.  

To elaborate on (ii) we have constructed flexible theories of the line shape that accurately and quantitatively account for both our numerical results and the experimental results: but only, paradoxically, under different limits.  Specifically, in experiment magnetic monopoles appear to screen the dipolar fields, leading to broad pinch points, while in simulation they do not screen, leading to sharp pinch points. 

In seeking a resolution of the pinch point paradox, we believe that a first obvious possibility --  pinch point broadening by structural defects and disorder in the samples -- may immediately be discarded. Although real samples do contain defects and disorder (see Refs.~\cite{Sala, Revell, Ghasemi} for a discussion of this and how to prepare defect free samples), theory appears to rule this out as a cause of pinch point broadening.  Thus, the role of defects and disorder has been considered theoretically in Ref.~\cite{Sen}. The case of non-magnetic dilution was shown to lead to no broadening of the dipolar pinch points.  Since our results agree with this theory in other respects, it therefore seems highly unlikely that a small concentration of defects and disorder in the experimental samples can give the very strong pinch point broadening that is actually observed. 

We are also confident that a second obvious possibility -- broadening by extra terms in the spin Hamiltonian -- may also be ruled out. We have checked this by simulating the refined $g-$DSI model (see Section II) which has been shown to give a most accurate description of ${\rm Dy_2Ti_2O_7}$~\cite{Yavorskii}. We find differences with DSI only near zone boundaries (see~\cref{highresfig}).  This is to be expected:  in this dipolar paramagnet,  only long-ranged (i.e dipolar) interactions should affect the correlation function at small $q$. We also tested the `dumbbell model'~\cite{CMS} (as implemented in Ref.~\cite{Bovo_special}), but found this to be qualitatively the same as DSI.

Similarly, we have considered the possibility of various experimental corrections, but none of these give a convincing explanation of the paradox. For example, incomplete beam polarization, corrections to the static approximation, or other sources of background would not lead to the systematics that we observe. The fact that neutron data taken under different conditions on different instruments and sources are in excellent agreement further points against experimental corrections being complicit. It is also crucial to emphasize, as discussed above, that Chang {\it et al.}'s analysis of their data 
(\cref{paradoxfiga}) is quantitatively consistent with ours: this is a crucial proof that there is no inadvertent bias in our analysis (for example a scaling of the experimental data onto one theoretical curve rather than the other in~\cref{paradoxfiga}).

These considerations leave us to face the likelihood of more fundamental causes of the pinch point paradox, including quantum fluctuations, corrections to neutron scattering theory and corrections to dipolar simulation methods. 

\medskip
\noindent
{\bf Quantum fluctuations -- } We have treated spin ice as a classical spin system. A careful appraisal of the problem~\cite{Gingras_quantum} showed that the experimental systems considered here should approximate classical dipolar spin ice very closely. The only correction would be at the monopole sites where there is experimental evidence of a second order Zeeman effect, indicating the superposition of `spin up' and `spin down' states associated with magnetic monopoles~\cite{Bovo,Paulsen_nuclear}. However, it is hard to see this having a significant effect on the correlation function or structure factor measured in the static approximation of neutron scattering.

\medskip
\noindent
{\bf Corrections to neutron scattering theory -- } The theory of neutron polarization analysis is well established following the pioneering work of Halpern and Holstein~\cite{Halpern} and Moon {\it et al.}~\cite{Moon}. It is customary to map the orbital contributions to the magnetic moment on to effective spin operators, but the experimental systems considered here have largely orbital, rather than spin magnetism. We suggest it would be worth closely examining the polarization analysis theory for spin ice. For example, it is certain that there would be some degree of interaction of the neutron `within' the orbital dipole, i.e. with the self-field, and we speculate that this might have some effect on the structure factor. Similarly, we might imagine some rather complex depolarization effects in spin ice, but these  would have to go well beyond the standard picture of depolarization to explain our observations. 

\medskip
\noindent
{\bf Corrections to dipolar simulation method -- } Clearly, the method used agrees with experiment at large $q$ and at $q = 0$, but does not do so at small but finite $q$. If we take the analytical theory to be a guide, we can expect the pinch point broadening to occur, regardless of the boundary conditions, so that it would even be present for periodic (Ewald) boundaries. Although it seems most unlikely that a tried and tested method like Ewald summation would have some basic flaw, it is also unlikely that it has ever been subject to such a stringent experimental test, so this question is worth examining in much greater detail.  In this context, it is worth emphasizing that the simulation does agree with experiment in so far as the total scattering or trace of correlation function: the discrepancy is at the level of the Helmholtz decomposition or polarization analysis. As above, it would be worth reconsidering the role of `self field' (the `internal structure of the dipole') as an obvious point of difference between theory and experiment. It is not present in simulation, the demagnetizing effect being implemented via a demagnetizing tensor, but it remains a very real and potentially important~\cite{Griffiths} experimental property.

\medskip
\noindent
{\bf Conclusion -- }
In conclusion, pinch points typify an important and widespread correlation in condensed matter. As we have stressed, they do not necessarily imply a Coulomb phase or even a singular structure factor, but they remain a key diagnostic of long ranged, or critical correlations in diverse systems. Topics of current theoretical interest include the coexistence of pinch points with ordered states~\cite{Banks, BrooksBartlett, Benton_frag},  extensions of the pinch point concept~\cite{Yan, Benton_pinchline},  pinch points as a diagnostic of quantum fluctuations~\cite{Benton, Benton_ice} and pinch points in Coulomb systems~\cite{McClarty, Slobinsky, Gray}.  In general a sharp pinch point indicates that the structure factor tensor remains eccentric as $q\rightarrow 0$, typically as as a result of unscreened power law correlations. These pinch points will broaden if the long range correlations are exponentially damped, as occurs with the screening of long ranged interactions. Using very high resolution dipolar simulations and by comparing to existing experiment, we have exposed an unusual paradox in dipolar spin ice: the dipolar fields are not screened in simulation but do appear to be screened in experiment. 

From a purely theoretical perspective, the pinch point paradox may be summarized as follows. In description of experiment, the Poisson-Boltzmann equation for magnetic monopoles appears where the Poisson equation is expected: it is as if the magnetic monopoles of spin ice behave much more like `real charges' (Dirac monopoles) than expected~\cite{CMS}.

More generally, given the special status of spin ice as a model ice rule system, a model dipolar system and a model Coulomb fluid, it is important to find the resolution of this paradox. When considered alongside recent results on ionic liquids~\cite{Lee} it seems that the screening of long ranged interactions in condensed matter presents many aspects that remain to be understood.

\begin{acknowledgments}
It is a pleasure to thank T. Fennell, P. C. W. Holdsworth and C. Gray for many useful discussions and related collaborations and P. C. W. Holdsworth for critical comments on the manuscript. The simulations were performed on resources provided by the Swedish National Infrastructure for Computing (SNIC) at the Center for High Performance Computing (PDC) at the Royal Institute of Technology (KTH). We gratefully acknowledge the NVIDIA Corporation for the donation of GPU resources. M.T. is supported by Stiftelsen Olle Engkvist Byggmästare (187-0013) with support from Magnus Bergvalls Stiftelse (2018-02701).  

The authors declare no competing financial interests.
\end{acknowledgments}


\begin{thebibliography}{83}%
\makeatletter
\providecommand \@ifxundefined [1]{%
 \@ifx{#1\undefined}
}%
\providecommand \@ifnum [1]{%
 \ifnum #1\expandafter \@firstoftwo
 \else \expandafter \@secondoftwo
 \fi
}%
\providecommand \@ifx [1]{%
 \ifx #1\expandafter \@firstoftwo
 \else \expandafter \@secondoftwo
 \fi
}%
\providecommand \natexlab [1]{#1}%
\providecommand \enquote  [1]{``#1''}%
\providecommand \bibnamefont  [1]{#1}%
\providecommand \bibfnamefont [1]{#1}%
\providecommand \citenamefont [1]{#1}%
\providecommand \href@noop [0]{\@secondoftwo}%
\providecommand \href [0]{\begingroup \@sanitize@url \@href}%
\providecommand \@href[1]{\@@startlink{#1}\@@href}%
\providecommand \@@href[1]{\endgroup#1\@@endlink}%
\providecommand \@sanitize@url [0]{\catcode `\\12\catcode `\$12\catcode
  `\&12\catcode `\#12\catcode `\^12\catcode `\_12\catcode `\%12\relax}%
\providecommand \@@startlink[1]{}%
\providecommand \@@endlink[0]{}%
\providecommand \url  [0]{\begingroup\@sanitize@url \@url }%
\providecommand \@url [1]{\endgroup\@href {#1}{\urlprefix }}%
\providecommand \urlprefix  [0]{URL }%
\providecommand \Eprint [0]{\href }%
\providecommand \doibase [0]{http://dx.doi.org/}%
\providecommand \selectlanguage [0]{\@gobble}%
\providecommand \bibinfo  [0]{\@secondoftwo}%
\providecommand \bibfield  [0]{\@secondoftwo}%
\providecommand \translation [1]{[#1]}%
\providecommand \BibitemOpen [0]{}%
\providecommand \bibitemStop [0]{}%
\providecommand \bibitemNoStop [0]{.\EOS\space}%
\providecommand \EOS [0]{\spacefactor3000\relax}%
\providecommand \BibitemShut  [1]{\csname bibitem#1\endcsname}%
\let\auto@bib@innerbib\@empty
\bibitem [{\citenamefont {Krivoglaz}(1964)}]{Krivoglaz}%
  \BibitemOpen
  \bibfield  {author} {\bibinfo {author} {\bibfnamefont {M.~A.}\ \bibnamefont
  {Krivoglaz}},\ }\bibfield  {title} {\enquote {\bibinfo {title} {Effect of
  long-range forces on fluctuations and the scattering of waves in crystals},}\
  }\href@noop {} {\bibfield  {journal} {\bibinfo  {journal} {Soviet
  Physics-Solid State}\ }\textbf {\bibinfo {volume} {5}},\ \bibinfo {pages}
  {2526--2534} (\bibinfo {year} {1964})}\BibitemShut {NoStop}%
\bibitem [{\citenamefont {Als-Nielsen}\ \emph {et~al.}(1976)\citenamefont
  {Als-Nielsen}, \citenamefont {Holmes},\ and\ \citenamefont
  {Guggenheim}}]{Als}%
  \BibitemOpen
  \bibfield  {author} {\bibinfo {author} {\bibfnamefont {J.}~\bibnamefont
  {Als-Nielsen}}, \bibinfo {author} {\bibfnamefont {L.~M.}\ \bibnamefont
  {Holmes}}, \ and\ \bibinfo {author} {\bibfnamefont {H.~J.}\ \bibnamefont
  {Guggenheim}},\ }\bibfield  {title} {\enquote {\bibinfo {title} {Experimental
  test of renormalization group theory on the uniaxial, dipolar coupled
  ferromagnet $\textup{LiTbF}_4$},}\ }\href {\doibase
  10.1103/PhysRevLett.37.1161} {\bibfield  {journal} {\bibinfo  {journal}
  {Phys. Rev. Lett.}\ }\textbf {\bibinfo {volume} {37}},\ \bibinfo {pages}
  {1161--1164} (\bibinfo {year} {1976})}\BibitemShut {NoStop}%
\bibitem [{\citenamefont {Paul}\ \emph {et~al.}(1970)\citenamefont {Paul},
  \citenamefont {Cochran}, \citenamefont {Buyers},\ and\ \citenamefont
  {Cowley}}]{Paul}%
  \BibitemOpen
  \bibfield  {author} {\bibinfo {author} {\bibfnamefont {G.~L.}\ \bibnamefont
  {Paul}}, \bibinfo {author} {\bibfnamefont {W.}~\bibnamefont {Cochran}},
  \bibinfo {author} {\bibfnamefont {W.~J.~L.}\ \bibnamefont {Buyers}}, \ and\
  \bibinfo {author} {\bibfnamefont {R.~A.}\ \bibnamefont {Cowley}},\ }\bibfield
   {title} {\enquote {\bibinfo {title} {Ferroelectric transition in
  $\textup{KD}_2\textup{PO}_4$},}\ }\href {\doibase 10.1103/PhysRevB.2.4603}
  {\bibfield  {journal} {\bibinfo  {journal} {Phys. Rev. B}\ }\textbf {\bibinfo
  {volume} {2}},\ \bibinfo {pages} {4603--4612} (\bibinfo {year}
  {1970})}\BibitemShut {NoStop}%
\bibitem [{\citenamefont {Havlin}\ \emph {et~al.}(1982)\citenamefont {Havlin},
  \citenamefont {Cowley},\ and\ \citenamefont {Sompolinsky}}]{Havlin}%
  \BibitemOpen
  \bibfield  {author} {\bibinfo {author} {\bibfnamefont {S.}~\bibnamefont
  {Havlin}}, \bibinfo {author} {\bibfnamefont {R.~A.}\ \bibnamefont {Cowley}},
  \ and\ \bibinfo {author} {\bibfnamefont {H.}~\bibnamefont {Sompolinsky}},\
  }\bibfield  {title} {\enquote {\bibinfo {title} {Anisotropic polarisation
  fluctuations in $\textup{KD}_2\textup{PO}_4$-type crystals},}\ }\href
  {\doibase 10.1088/0022-3719/15/29/017} {\bibfield  {journal} {\bibinfo
  {journal} {Journal of Physics C: Solid State Physics}\ }\textbf {\bibinfo
  {volume} {15}},\ \bibinfo {pages} {6057--6066} (\bibinfo {year}
  {1982})}\BibitemShut {NoStop}%
\bibitem [{\citenamefont {Youngblood}\ and\ \citenamefont {Axe}(1978)}]{YA}%
  \BibitemOpen
  \bibfield  {author} {\bibinfo {author} {\bibfnamefont {R.}~\bibnamefont
  {Youngblood}}\ and\ \bibinfo {author} {\bibfnamefont {J.~D.}\ \bibnamefont
  {Axe}},\ }\bibfield  {title} {\enquote {\bibinfo {title} {Neutron-scattering
  study of short-range order in a model two-dimensional ferroelectric},}\
  }\href {\doibase 10.1103/PhysRevB.17.3639} {\bibfield  {journal} {\bibinfo
  {journal} {Phys. Rev. B}\ }\textbf {\bibinfo {volume} {17}},\ \bibinfo
  {pages} {3639--3649} (\bibinfo {year} {1978})}\BibitemShut {NoStop}%
\bibitem [{\citenamefont {Youngblood}\ \emph {et~al.}(1980)\citenamefont
  {Youngblood}, \citenamefont {Axe},\ and\ \citenamefont {McCoy}}]{YAM}%
  \BibitemOpen
  \bibfield  {author} {\bibinfo {author} {\bibfnamefont {R.}~\bibnamefont
  {Youngblood}}, \bibinfo {author} {\bibfnamefont {J.~D.}\ \bibnamefont {Axe}},
  \ and\ \bibinfo {author} {\bibfnamefont {B.~M.}\ \bibnamefont {McCoy}},\
  }\bibfield  {title} {\enquote {\bibinfo {title} {Correlations in ice-rule
  ferroelectrics},}\ }\href {\doibase 10.1103/PhysRevB.21.5212} {\bibfield
  {journal} {\bibinfo  {journal} {Phys. Rev. B}\ }\textbf {\bibinfo {volume}
  {21}},\ \bibinfo {pages} {5212--5220} (\bibinfo {year} {1980})}\BibitemShut
  {NoStop}%
\bibitem [{\citenamefont {Nield}\ and\ \citenamefont
  {Whitworth}(1995)}]{Nield}%
  \BibitemOpen
  \bibfield  {author} {\bibinfo {author} {\bibfnamefont {V.~M.}\ \bibnamefont
  {Nield}}\ and\ \bibinfo {author} {\bibfnamefont {R.~W.}\ \bibnamefont
  {Whitworth}},\ }\bibfield  {title} {\enquote {\bibinfo {title} {The structure
  of ice $\textup{I}_{\textup{h}}$ from analysis of single-crystal neutron
  diffuse scattering},}\ }\href {\doibase 10.1088/0953-8984/7/43/006}
  {\bibfield  {journal} {\bibinfo  {journal} {Journal of Physics: Condensed
  Matter}\ }\textbf {\bibinfo {volume} {7}},\ \bibinfo {pages} {8259--8271}
  (\bibinfo {year} {1995})}\BibitemShut {NoStop}%
\bibitem [{\citenamefont {Wehinger}\ \emph {et~al.}(2014)\citenamefont
  {Wehinger}, \citenamefont {Chernyshov}, \citenamefont {Krisch}, \citenamefont
  {Bulat}, \citenamefont {Ezhov},\ and\ \citenamefont {Bosak}}]{Wehinger}%
  \BibitemOpen
  \bibfield  {author} {\bibinfo {author} {\bibfnamefont {B.}~\bibnamefont
  {Wehinger}}, \bibinfo {author} {\bibfnamefont {D.}~\bibnamefont
  {Chernyshov}}, \bibinfo {author} {\bibfnamefont {M.}~\bibnamefont {Krisch}},
  \bibinfo {author} {\bibfnamefont {S.}~\bibnamefont {Bulat}}, \bibinfo
  {author} {\bibfnamefont {V.}~\bibnamefont {Ezhov}}, \ and\ \bibinfo {author}
  {\bibfnamefont {A.}~\bibnamefont {Bosak}},\ }\bibfield  {title} {\enquote
  {\bibinfo {title} {Diffuse scattering in $\textup{I}_{\textup{h}}$ ice},}\
  }\href {\doibase 10.1088/0953-8984/26/26/265401} {\bibfield  {journal}
  {\bibinfo  {journal} {Journal of Physics: Condensed Matter}\ }\textbf
  {\bibinfo {volume} {26}},\ \bibinfo {pages} {265401} (\bibinfo {year}
  {2014})}\BibitemShut {NoStop}%
\bibitem [{\citenamefont {Moessner}\ and\ \citenamefont {Sondhi}(2003)}]{MS}%
  \BibitemOpen
  \bibfield  {author} {\bibinfo {author} {\bibfnamefont {R.}~\bibnamefont
  {Moessner}}\ and\ \bibinfo {author} {\bibfnamefont {S.~L.}\ \bibnamefont
  {Sondhi}},\ }\bibfield  {title} {\enquote {\bibinfo {title} {Theory of the
  [111] magnetization plateau in spin ice},}\ }\href {\doibase
  10.1103/PhysRevB.68.064411} {\bibfield  {journal} {\bibinfo  {journal} {Phys.
  Rev. B}\ }\textbf {\bibinfo {volume} {68}},\ \bibinfo {pages} {064411}
  (\bibinfo {year} {2003})}\BibitemShut {NoStop}%
\bibitem [{\citenamefont {Fennell}\ \emph {et~al.}(2007)\citenamefont
  {Fennell}, \citenamefont {Bramwell}, \citenamefont {McMorrow}, \citenamefont
  {Manuel},\ and\ \citenamefont {Wildes}}]{Fennell_kag}%
  \BibitemOpen
  \bibfield  {author} {\bibinfo {author} {\bibfnamefont {T.}~\bibnamefont
  {Fennell}}, \bibinfo {author} {\bibfnamefont {S.~T.}\ \bibnamefont
  {Bramwell}}, \bibinfo {author} {\bibfnamefont {D.~F.}\ \bibnamefont
  {McMorrow}}, \bibinfo {author} {\bibfnamefont {P.}~\bibnamefont {Manuel}}, \
  and\ \bibinfo {author} {\bibfnamefont {A.~R.}\ \bibnamefont {Wildes}},\
  }\bibfield  {title} {\enquote {\bibinfo {title} {Pinch points and
  $\textup{K}$asteleyn transitions in kagome ice},}\ }\href
  {https://doi.org/10.1038/nphys632} {\bibfield  {journal} {\bibinfo  {journal}
  {Nature Physics}\ }\textbf {\bibinfo {volume} {3}},\ \bibinfo {pages} {566}
  (\bibinfo {year} {2007})}\BibitemShut {NoStop}%
\bibitem [{\citenamefont {Fennell}\ \emph {et~al.}(2009)\citenamefont
  {Fennell}, \citenamefont {Deen}, \citenamefont {Wildes}, \citenamefont
  {Schmalzl}, \citenamefont {Prabhakaran}, \citenamefont {Boothroyd},
  \citenamefont {Aldus}, \citenamefont {McMorrow},\ and\ \citenamefont
  {Bramwell}}]{Fennell}%
  \BibitemOpen
  \bibfield  {author} {\bibinfo {author} {\bibfnamefont {T.}~\bibnamefont
  {Fennell}}, \bibinfo {author} {\bibfnamefont {P.~P.}\ \bibnamefont {Deen}},
  \bibinfo {author} {\bibfnamefont {A.~R.}\ \bibnamefont {Wildes}}, \bibinfo
  {author} {\bibfnamefont {K.}~\bibnamefont {Schmalzl}}, \bibinfo {author}
  {\bibfnamefont {D.}~\bibnamefont {Prabhakaran}}, \bibinfo {author}
  {\bibfnamefont {A.~T.}\ \bibnamefont {Boothroyd}}, \bibinfo {author}
  {\bibfnamefont {R.~J.}\ \bibnamefont {Aldus}}, \bibinfo {author}
  {\bibfnamefont {D.~F.}\ \bibnamefont {McMorrow}}, \ and\ \bibinfo {author}
  {\bibfnamefont {S.~T.}\ \bibnamefont {Bramwell}},\ }\bibfield  {title}
  {\enquote {\bibinfo {title} {Magnetic $\textup{C}$oulomb phase in the spin
  ice $\textup{Ho}_2\textup{Ti}_2\textup{O}_7$},}\ }\href {\doibase
  10.1126/science.1177582} {\bibfield  {journal} {\bibinfo  {journal}
  {Science}\ }\textbf {\bibinfo {volume} {326}},\ \bibinfo {pages} {415--417}
  (\bibinfo {year} {2009})}\BibitemShut {NoStop}%
\bibitem [{\citenamefont {Chang}\ \emph {et~al.}(2010)\citenamefont {Chang},
  \citenamefont {Su}, \citenamefont {Kao}, \citenamefont {Chou}, \citenamefont
  {Mittal}, \citenamefont {Schneider}, \citenamefont {Br\"uckel}, \citenamefont
  {Balakrishnan},\ and\ \citenamefont {Lees}}]{Chang}%
  \BibitemOpen
  \bibfield  {author} {\bibinfo {author} {\bibfnamefont {L.~J.}\ \bibnamefont
  {Chang}}, \bibinfo {author} {\bibfnamefont {Y.}~\bibnamefont {Su}}, \bibinfo
  {author} {\bibfnamefont {Y.-J.}\ \bibnamefont {Kao}}, \bibinfo {author}
  {\bibfnamefont {Y.~Z.}\ \bibnamefont {Chou}}, \bibinfo {author}
  {\bibfnamefont {R.}~\bibnamefont {Mittal}}, \bibinfo {author} {\bibfnamefont
  {H.}~\bibnamefont {Schneider}}, \bibinfo {author} {\bibfnamefont
  {T.}~\bibnamefont {Br\"uckel}}, \bibinfo {author} {\bibfnamefont
  {G.}~\bibnamefont {Balakrishnan}}, \ and\ \bibinfo {author} {\bibfnamefont
  {M.~R.}\ \bibnamefont {Lees}},\ }\bibfield  {title} {\enquote {\bibinfo
  {title} {Magnetic correlations in the spin ice
  $\textup{Ho}_{2\ensuremath{-}x}\textup{Y}_x\textup{Ti}_2\textup{O}_7$ as
  revealed by neutron polarization analysis},}\ }\href {\doibase
  10.1103/PhysRevB.82.172403} {\bibfield  {journal} {\bibinfo  {journal} {Phys.
  Rev. B}\ }\textbf {\bibinfo {volume} {82}},\ \bibinfo {pages} {172403}
  (\bibinfo {year} {2010})}\BibitemShut {NoStop}%
\bibitem [{\citenamefont {Sen}\ \emph {et~al.}(2013)\citenamefont {Sen},
  \citenamefont {Moessner},\ and\ \citenamefont {Sondhi}}]{Sen}%
  \BibitemOpen
  \bibfield  {author} {\bibinfo {author} {\bibfnamefont {A.}~\bibnamefont
  {Sen}}, \bibinfo {author} {\bibfnamefont {R.}~\bibnamefont {Moessner}}, \
  and\ \bibinfo {author} {\bibfnamefont {S.~L.}\ \bibnamefont {Sondhi}},\
  }\bibfield  {title} {\enquote {\bibinfo {title} {$\textup{C}$oulomb phase
  diagnostics as a function of temperature, interaction range, and disorder},}\
  }\href {\doibase 10.1103/PhysRevLett.110.107202} {\bibfield  {journal}
  {\bibinfo  {journal} {Phys. Rev. Lett.}\ }\textbf {\bibinfo {volume} {110}},\
  \bibinfo {pages} {107202} (\bibinfo {year} {2013})}\BibitemShut {NoStop}%
\bibitem [{\citenamefont {Ryzhkin}\ \emph {et~al.}(2013)\citenamefont
  {Ryzhkin}, \citenamefont {Ryzhkin},\ and\ \citenamefont {Bramwell}}]{RRB}%
  \BibitemOpen
  \bibfield  {author} {\bibinfo {author} {\bibfnamefont {M.~I.}\ \bibnamefont
  {Ryzhkin}}, \bibinfo {author} {\bibfnamefont {I.~A.}\ \bibnamefont
  {Ryzhkin}}, \ and\ \bibinfo {author} {\bibfnamefont {S.~T.}\ \bibnamefont
  {Bramwell}},\ }\bibfield  {title} {\enquote {\bibinfo {title} {Dynamic
  susceptibility and dynamic correlations in spin ice},}\ }\href {\doibase
  10.1209/0295-5075/104/37005} {\bibfield  {journal} {\bibinfo  {journal}
  {{EPL} (Europhysics Letters)}\ }\textbf {\bibinfo {volume} {104}},\ \bibinfo
  {pages} {37005} (\bibinfo {year} {2013})}\BibitemShut {NoStop}%
\bibitem [{\citenamefont {Anderson}(1956)}]{Anderson}%
  \BibitemOpen
  \bibfield  {author} {\bibinfo {author} {\bibfnamefont {P.~W.}\ \bibnamefont
  {Anderson}},\ }\bibfield  {title} {\enquote {\bibinfo {title} {Ordering and
  antiferromagnetism in ferrites},}\ }\href {\doibase 10.1103/PhysRev.102.1008}
  {\bibfield  {journal} {\bibinfo  {journal} {Phys. Rev.}\ }\textbf {\bibinfo
  {volume} {102}},\ \bibinfo {pages} {1008--1013} (\bibinfo {year}
  {1956})}\BibitemShut {NoStop}%
\bibitem [{\citenamefont {Fennell}\ \emph {et~al.}(2019)\citenamefont
  {Fennell}, \citenamefont {Harris}, \citenamefont {Calder}, \citenamefont
  {Ruminy}, \citenamefont {Boehm}, \citenamefont {Steffens}, \citenamefont
  {Lemee-Cailleau}, \citenamefont {Zaharko}, \citenamefont {Cervellino},\ and\
  \citenamefont {Bramwell}}]{Fennell2019}%
  \BibitemOpen
  \bibfield  {author} {\bibinfo {author} {\bibfnamefont {T.}~\bibnamefont
  {Fennell}}, \bibinfo {author} {\bibfnamefont {M.}~\bibnamefont {Harris}},
  \bibinfo {author} {\bibfnamefont {S.}~\bibnamefont {Calder}}, \bibinfo
  {author} {\bibfnamefont {M.}~\bibnamefont {Ruminy}}, \bibinfo {author}
  {\bibfnamefont {M.}~\bibnamefont {Boehm}}, \bibinfo {author} {\bibfnamefont
  {P.}~\bibnamefont {Steffens}}, \bibinfo {author} {\bibfnamefont {M.~H.}\
  \bibnamefont {Lemee-Cailleau}}, \bibinfo {author} {\bibfnamefont
  {O.}~\bibnamefont {Zaharko}}, \bibinfo {author} {\bibfnamefont
  {A.}~\bibnamefont {Cervellino}}, \ and\ \bibinfo {author} {\bibfnamefont
  {S.~T.}\ \bibnamefont {Bramwell}},\ }\bibfield  {title} {\enquote {\bibinfo
  {title} {Multiple $\textup{C}$oulomb phase in the fluoride pyrochlore
  $\textup{CsNiCrF}_6$},}\ }\href {\doibase 10.1038/s41567-018-0309-3}
  {\bibfield  {journal} {\bibinfo  {journal} {Nature Physics}\ }\textbf
  {\bibinfo {volume} {15}},\ \bibinfo {pages} {60--66} (\bibinfo {year}
  {2019})}\BibitemShut {NoStop}%
\bibitem [{\citenamefont {Perrin}\ \emph {et~al.}(2016)\citenamefont {Perrin},
  \citenamefont {Canals},\ and\ \citenamefont {Rougemaille}}]{Perrin}%
  \BibitemOpen
  \bibfield  {author} {\bibinfo {author} {\bibfnamefont {Y.}~\bibnamefont
  {Perrin}}, \bibinfo {author} {\bibfnamefont {B.}~\bibnamefont {Canals}}, \
  and\ \bibinfo {author} {\bibfnamefont {N.}~\bibnamefont {Rougemaille}},\
  }\bibfield  {title} {\enquote {\bibinfo {title} {Extensive degeneracy,
  $\textup{C}$oulomb phase and magnetic monopoles in artificial square ice},}\
  }\href {https://doi.org/10.1038/nature20155} {\bibfield  {journal} {\bibinfo
  {journal} {Nature}\ }\textbf {\bibinfo {volume} {540}},\ \bibinfo {pages}
  {410} (\bibinfo {year} {2016})}\BibitemShut {NoStop}%
\bibitem [{\citenamefont {{\"O}stman}\ \emph {et~al.}(2018)\citenamefont
  {{\"O}stman}, \citenamefont {Stopfel}, \citenamefont {Chioar}, \citenamefont
  {Arnalds}, \citenamefont {Stein}, \citenamefont {Kapaklis},\ and\
  \citenamefont {Hj{\"o}rvarsson}}]{Ostman}%
  \BibitemOpen
  \bibfield  {author} {\bibinfo {author} {\bibfnamefont {E.}~\bibnamefont
  {{\"O}stman}}, \bibinfo {author} {\bibfnamefont {H.}~\bibnamefont {Stopfel}},
  \bibinfo {author} {\bibfnamefont {I.-A.}\ \bibnamefont {Chioar}}, \bibinfo
  {author} {\bibfnamefont {U.~B.}\ \bibnamefont {Arnalds}}, \bibinfo {author}
  {\bibfnamefont {A.}~\bibnamefont {Stein}}, \bibinfo {author} {\bibfnamefont
  {V.}~\bibnamefont {Kapaklis}}, \ and\ \bibinfo {author} {\bibfnamefont
  {B.}~\bibnamefont {Hj{\"o}rvarsson}},\ }\bibfield  {title} {\enquote
  {\bibinfo {title} {Interaction modifiers in artificial spin ices},}\ }\href
  {\doibase 10.1038/s41567-017-0027-2} {\bibfield  {journal} {\bibinfo
  {journal} {Nature Physics}\ }\textbf {\bibinfo {volume} {14}},\ \bibinfo
  {pages} {375--379} (\bibinfo {year} {2018})}\BibitemShut {NoStop}%
\bibitem [{\citenamefont {Farhan}\ \emph {et~al.}(2019)\citenamefont {Farhan},
  \citenamefont {Saccone}, \citenamefont {Petersen}, \citenamefont {Dhuey},
  \citenamefont {Chopdekar}, \citenamefont {Huang}, \citenamefont {Kent},
  \citenamefont {Chen}, \citenamefont {Alava}, \citenamefont {Lippert},
  \citenamefont {Scholl},\ and\ \citenamefont {van Dijken}}]{Farhan}%
  \BibitemOpen
  \bibfield  {author} {\bibinfo {author} {\bibfnamefont {A.}~\bibnamefont
  {Farhan}}, \bibinfo {author} {\bibfnamefont {M.}~\bibnamefont {Saccone}},
  \bibinfo {author} {\bibfnamefont {C.~F.}\ \bibnamefont {Petersen}}, \bibinfo
  {author} {\bibfnamefont {S.}~\bibnamefont {Dhuey}}, \bibinfo {author}
  {\bibfnamefont {R.~V.}\ \bibnamefont {Chopdekar}}, \bibinfo {author}
  {\bibfnamefont {Y.-L.}\ \bibnamefont {Huang}}, \bibinfo {author}
  {\bibfnamefont {N.}~\bibnamefont {Kent}}, \bibinfo {author} {\bibfnamefont
  {Z.}~\bibnamefont {Chen}}, \bibinfo {author} {\bibfnamefont {M.~J.}\
  \bibnamefont {Alava}}, \bibinfo {author} {\bibfnamefont {T.}~\bibnamefont
  {Lippert}}, \bibinfo {author} {\bibfnamefont {A.}~\bibnamefont {Scholl}}, \
  and\ \bibinfo {author} {\bibfnamefont {S.}~\bibnamefont {van Dijken}},\
  }\bibfield  {title} {\enquote {\bibinfo {title} {Emergent magnetic monopole
  dynamics in macroscopically degenerate artificial spin ice},}\ }\href
  {https://advances.sciencemag.org/content/5/2/eaav6380} {\bibfield  {journal}
  {\bibinfo  {journal} {Science Advances}\ }\textbf {\bibinfo {volume} {5}}
  (\bibinfo {year} {2019})}\BibitemShut {NoStop}%
\bibitem [{\citenamefont {Benton}\ \emph {et~al.}(2012)\citenamefont {Benton},
  \citenamefont {Sikora},\ and\ \citenamefont {Shannon}}]{Benton}%
  \BibitemOpen
  \bibfield  {author} {\bibinfo {author} {\bibfnamefont {O.}~\bibnamefont
  {Benton}}, \bibinfo {author} {\bibfnamefont {O.}~\bibnamefont {Sikora}}, \
  and\ \bibinfo {author} {\bibfnamefont {N.}~\bibnamefont {Shannon}},\
  }\bibfield  {title} {\enquote {\bibinfo {title} {Seeing the light:
  Experimental signatures of emergent electromagnetism in a quantum spin
  ice},}\ }\href {\doibase 10.1103/PhysRevB.86.075154} {\bibfield  {journal}
  {\bibinfo  {journal} {Phys. Rev. B}\ }\textbf {\bibinfo {volume} {86}},\
  \bibinfo {pages} {075154} (\bibinfo {year} {2012})}\BibitemShut {NoStop}%
\bibitem [{\citenamefont {Fennell}\ \emph {et~al.}(2012)\citenamefont
  {Fennell}, \citenamefont {Kenzelmann}, \citenamefont {Roessli}, \citenamefont
  {Haas},\ and\ \citenamefont {Cava}}]{Fennell_TTO}%
  \BibitemOpen
  \bibfield  {author} {\bibinfo {author} {\bibfnamefont {T.}~\bibnamefont
  {Fennell}}, \bibinfo {author} {\bibfnamefont {M.}~\bibnamefont {Kenzelmann}},
  \bibinfo {author} {\bibfnamefont {B.}~\bibnamefont {Roessli}}, \bibinfo
  {author} {\bibfnamefont {M.~K.}\ \bibnamefont {Haas}}, \ and\ \bibinfo
  {author} {\bibfnamefont {R.~J.}\ \bibnamefont {Cava}},\ }\bibfield  {title}
  {\enquote {\bibinfo {title} {Power-law spin correlations in the pyrochlore
  antiferromagnet $\textup{Tb}_2\textup{Ti}_2\textup{O}_7$},}\ }\href {\doibase
  10.1103/PhysRevLett.109.017201} {\bibfield  {journal} {\bibinfo  {journal}
  {Phys. Rev. Lett.}\ }\textbf {\bibinfo {volume} {109}},\ \bibinfo {pages}
  {017201} (\bibinfo {year} {2012})}\BibitemShut {NoStop}%
\bibitem [{\citenamefont {Sibille}\ \emph {et~al.}(2018)\citenamefont
  {Sibille}, \citenamefont {Gauthier}, \citenamefont {Yan}, \citenamefont
  {Ciomaga~Hatnean}, \citenamefont {Ollivier}, \citenamefont {Winn},
  \citenamefont {Filges}, \citenamefont {Balakrishnan}, \citenamefont
  {Kenzelmann}, \citenamefont {Shannon},\ and\ \citenamefont
  {Fennell}}]{Sibille}%
  \BibitemOpen
  \bibfield  {author} {\bibinfo {author} {\bibfnamefont {R.}~\bibnamefont
  {Sibille}}, \bibinfo {author} {\bibfnamefont {N.}~\bibnamefont {Gauthier}},
  \bibinfo {author} {\bibfnamefont {H.}~\bibnamefont {Yan}}, \bibinfo {author}
  {\bibfnamefont {M.}~\bibnamefont {Ciomaga~Hatnean}}, \bibinfo {author}
  {\bibfnamefont {J.}~\bibnamefont {Ollivier}}, \bibinfo {author}
  {\bibfnamefont {B.}~\bibnamefont {Winn}}, \bibinfo {author} {\bibfnamefont
  {U.}~\bibnamefont {Filges}}, \bibinfo {author} {\bibfnamefont
  {G.}~\bibnamefont {Balakrishnan}}, \bibinfo {author} {\bibfnamefont
  {M.}~\bibnamefont {Kenzelmann}}, \bibinfo {author} {\bibfnamefont
  {N.}~\bibnamefont {Shannon}}, \ and\ \bibinfo {author} {\bibfnamefont
  {T.}~\bibnamefont {Fennell}},\ }\bibfield  {title} {\enquote {\bibinfo
  {title} {Experimental signatures of emergent quantum electrodynamics in
  $\textup{Pr}_2\textup{Hf}_2\textup{O}_7$},}\ }\href {\doibase
  10.1038/s41567-018-0116-x} {\bibfield  {journal} {\bibinfo  {journal} {Nature
  Physics}\ }\textbf {\bibinfo {volume} {14}},\ \bibinfo {pages} {711--715}
  (\bibinfo {year} {2018})}\BibitemShut {NoStop}%
\bibitem [{\citenamefont {Zinkin}\ \emph {et~al.}(1997)\citenamefont {Zinkin},
  \citenamefont {Harris},\ and\ \citenamefont {Zeiske}}]{Zinkin}%
  \BibitemOpen
  \bibfield  {author} {\bibinfo {author} {\bibfnamefont {M.~P.}\ \bibnamefont
  {Zinkin}}, \bibinfo {author} {\bibfnamefont {M.~J.}\ \bibnamefont {Harris}},
  \ and\ \bibinfo {author} {\bibfnamefont {T.}~\bibnamefont {Zeiske}},\
  }\bibfield  {title} {\enquote {\bibinfo {title} {Short-range magnetic order
  in the frustrated pyrochlore antiferromagnet $\textup{CsNiCrF}_{6}$},}\
  }\href {\doibase 10.1103/PhysRevB.56.11786} {\bibfield  {journal} {\bibinfo
  {journal} {Phys. Rev. B}\ }\textbf {\bibinfo {volume} {56}},\ \bibinfo
  {pages} {11786--11790} (\bibinfo {year} {1997})}\BibitemShut {NoStop}%
\bibitem [{\citenamefont {Conlon}\ and\ \citenamefont
  {Chalker}(2009)}]{Conlon}%
  \BibitemOpen
  \bibfield  {author} {\bibinfo {author} {\bibfnamefont {P.~H.}\ \bibnamefont
  {Conlon}}\ and\ \bibinfo {author} {\bibfnamefont {J.~T.}\ \bibnamefont
  {Chalker}},\ }\bibfield  {title} {\enquote {\bibinfo {title} {Spin dynamics
  in pyrochlore $\textup{H}$eisenberg antiferromagnets},}\ }\href {\doibase
  10.1103/PhysRevLett.102.237206} {\bibfield  {journal} {\bibinfo  {journal}
  {Phys. Rev. Lett.}\ }\textbf {\bibinfo {volume} {102}},\ \bibinfo {pages}
  {237206} (\bibinfo {year} {2009})}\BibitemShut {NoStop}%
\bibitem [{\citenamefont {Ballou}\ \emph {et~al.}(1996)\citenamefont {Ballou},
  \citenamefont {Leli\`evre-Berna},\ and\ \citenamefont {F\aa{}k}}]{Ballou}%
  \BibitemOpen
  \bibfield  {author} {\bibinfo {author} {\bibfnamefont {R.}~\bibnamefont
  {Ballou}}, \bibinfo {author} {\bibfnamefont {E.}~\bibnamefont
  {Leli\`evre-Berna}}, \ and\ \bibinfo {author} {\bibfnamefont
  {B.}~\bibnamefont {F\aa{}k}},\ }\bibfield  {title} {\enquote {\bibinfo
  {title} {Spin fluctuations in
  $(\textup{Y}_{0.97}\textup{Sc}_{0.03})\textup{Mn}_{2}$: A geometrically
  frustrated, nearly antiferromagnetic, itinerant electron system},}\ }\href
  {\doibase 10.1103/PhysRevLett.76.2125} {\bibfield  {journal} {\bibinfo
  {journal} {Phys. Rev. Lett.}\ }\textbf {\bibinfo {volume} {76}},\ \bibinfo
  {pages} {2125--2128} (\bibinfo {year} {1996})}\BibitemShut {NoStop}%
\bibitem [{\citenamefont {Canals}\ and\ \citenamefont
  {Lacroix}(2000)}]{Canals}%
  \BibitemOpen
  \bibfield  {author} {\bibinfo {author} {\bibfnamefont {B.}~\bibnamefont
  {Canals}}\ and\ \bibinfo {author} {\bibfnamefont {C.}~\bibnamefont
  {Lacroix}},\ }\bibfield  {title} {\enquote {\bibinfo {title} {Quantum spin
  liquid: The $\textup{H}$eisenberg antiferromagnet on the three-dimensional
  pyrochlore lattice},}\ }\href {\doibase 10.1103/PhysRevB.61.1149} {\bibfield
  {journal} {\bibinfo  {journal} {Phys. Rev. B}\ }\textbf {\bibinfo {volume}
  {61}},\ \bibinfo {pages} {1149--1159} (\bibinfo {year} {2000})}\BibitemShut
  {NoStop}%
\bibitem [{\citenamefont {Henley}(2005)}]{Henley}%
  \BibitemOpen
  \bibfield  {author} {\bibinfo {author} {\bibfnamefont {C.~L.}\ \bibnamefont
  {Henley}},\ }\bibfield  {title} {\enquote {\bibinfo {title} {Power-law spin
  correlations in pyrochlore antiferromagnets},}\ }\href {\doibase
  10.1103/PhysRevB.71.014424} {\bibfield  {journal} {\bibinfo  {journal} {Phys.
  Rev. B}\ }\textbf {\bibinfo {volume} {71}},\ \bibinfo {pages} {014424}
  (\bibinfo {year} {2005})}\BibitemShut {NoStop}%
\bibitem [{\citenamefont {Pauling}(1935)}]{Pauling}%
  \BibitemOpen
  \bibfield  {author} {\bibinfo {author} {\bibfnamefont {L.}~\bibnamefont
  {Pauling}},\ }\bibfield  {title} {\enquote {\bibinfo {title} {The structure
  and entropy of ice and of other crystals with some randomness of atomic
  arrangement},}\ }\href {\doibase 10.1021/ja01315a102} {\bibfield  {journal}
  {\bibinfo  {journal} {Journal of the American Chemical Society}\ }\textbf
  {\bibinfo {volume} {57}},\ \bibinfo {pages} {2680--2684} (\bibinfo {year}
  {1935})}\BibitemShut {NoStop}%
\bibitem [{\citenamefont {Ramirez}\ \emph {et~al.}(1999)\citenamefont
  {Ramirez}, \citenamefont {Hayashi}, \citenamefont {Cava}, \citenamefont
  {Siddharthan},\ and\ \citenamefont {Shastry}}]{Ramirez}%
  \BibitemOpen
  \bibfield  {author} {\bibinfo {author} {\bibfnamefont {A.~P.}\ \bibnamefont
  {Ramirez}}, \bibinfo {author} {\bibfnamefont {A.}~\bibnamefont {Hayashi}},
  \bibinfo {author} {\bibfnamefont {R.~J.}\ \bibnamefont {Cava}}, \bibinfo
  {author} {\bibfnamefont {R.}~\bibnamefont {Siddharthan}}, \ and\ \bibinfo
  {author} {\bibfnamefont {B.~S.}\ \bibnamefont {Shastry}},\ }\bibfield
  {title} {\enquote {\bibinfo {title} {Zero-point entropy in 'spin ice'},}\
  }\href {http://dx.doi.org/10.1038/20619} {\bibfield  {journal} {\bibinfo
  {journal} {Nature}\ }\textbf {\bibinfo {volume} {399}},\ \bibinfo {pages}
  {333 EP --} (\bibinfo {year} {1999})}\BibitemShut {NoStop}%
\bibitem [{\citenamefont {Giblin}\ \emph {et~al.}(2018)\citenamefont {Giblin},
  \citenamefont {Twengstr\"om}, \citenamefont {Bovo}, \citenamefont {Ruminy},
  \citenamefont {Bartkowiak}, \citenamefont {Manuel}, \citenamefont {Andresen},
  \citenamefont {Prabhakaran}, \citenamefont {Balakrishnan}, \citenamefont
  {Pomjakushina}, \citenamefont {Paulsen}, \citenamefont {Lhotel},
  \citenamefont {Keller}, \citenamefont {Frontzek}, \citenamefont {Capelli},
  \citenamefont {Zaharko}, \citenamefont {McClarty}, \citenamefont {Bramwell},
  \citenamefont {Henelius},\ and\ \citenamefont {Fennell}}]{Giblin}%
  \BibitemOpen
  \bibfield  {author} {\bibinfo {author} {\bibfnamefont {S.~R.}\ \bibnamefont
  {Giblin}}, \bibinfo {author} {\bibfnamefont {M.}~\bibnamefont
  {Twengstr\"om}}, \bibinfo {author} {\bibfnamefont {L.}~\bibnamefont {Bovo}},
  \bibinfo {author} {\bibfnamefont {M.}~\bibnamefont {Ruminy}}, \bibinfo
  {author} {\bibfnamefont {M.}~\bibnamefont {Bartkowiak}}, \bibinfo {author}
  {\bibfnamefont {P.}~\bibnamefont {Manuel}}, \bibinfo {author} {\bibfnamefont
  {J.~C.}\ \bibnamefont {Andresen}}, \bibinfo {author} {\bibfnamefont
  {D.}~\bibnamefont {Prabhakaran}}, \bibinfo {author} {\bibfnamefont
  {G.}~\bibnamefont {Balakrishnan}}, \bibinfo {author} {\bibfnamefont
  {E.}~\bibnamefont {Pomjakushina}}, \bibinfo {author} {\bibfnamefont
  {C.}~\bibnamefont {Paulsen}}, \bibinfo {author} {\bibfnamefont
  {E.}~\bibnamefont {Lhotel}}, \bibinfo {author} {\bibfnamefont
  {L.}~\bibnamefont {Keller}}, \bibinfo {author} {\bibfnamefont
  {M.}~\bibnamefont {Frontzek}}, \bibinfo {author} {\bibfnamefont {S.~C.}\
  \bibnamefont {Capelli}}, \bibinfo {author} {\bibfnamefont {O.}~\bibnamefont
  {Zaharko}}, \bibinfo {author} {\bibfnamefont {P.~A.}\ \bibnamefont
  {McClarty}}, \bibinfo {author} {\bibfnamefont {S.~T.}\ \bibnamefont
  {Bramwell}}, \bibinfo {author} {\bibfnamefont {P.}~\bibnamefont {Henelius}},
  \ and\ \bibinfo {author} {\bibfnamefont {T.}~\bibnamefont {Fennell}},\
  }\bibfield  {title} {\enquote {\bibinfo {title} {Pauling entropy,
  metastability, and equilibrium in $\textup{Dy}_2\textup{Ti}_2\textup{O}_7$
  spin ice},}\ }\href {\doibase 10.1103/PhysRevLett.121.067202} {\bibfield
  {journal} {\bibinfo  {journal} {Phys. Rev. Lett.}\ }\textbf {\bibinfo
  {volume} {121}},\ \bibinfo {pages} {067202} (\bibinfo {year}
  {2018})}\BibitemShut {NoStop}%
\bibitem [{\citenamefont {Bramwell}\ and\ \citenamefont
  {Gingras}(2001)}]{Bramwell_Gingras}%
  \BibitemOpen
  \bibfield  {author} {\bibinfo {author} {\bibfnamefont {S.~T.}\ \bibnamefont
  {Bramwell}}\ and\ \bibinfo {author} {\bibfnamefont {M.~J.~P.}\ \bibnamefont
  {Gingras}},\ }\bibfield  {title} {\enquote {\bibinfo {title} {Spin ice state
  in frustrated magnetic pyrochlore materials},}\ }\href {\doibase
  10.1126/science.1064761} {\bibfield  {journal} {\bibinfo  {journal}
  {Science}\ }\textbf {\bibinfo {volume} {294}},\ \bibinfo {pages} {1495--1501}
  (\bibinfo {year} {2001})}\BibitemShut {NoStop}%
\bibitem [{\citenamefont {Harris}\ \emph {et~al.}(1997)\citenamefont {Harris},
  \citenamefont {Bramwell}, \citenamefont {McMorrow}, \citenamefont {Zeiske},\
  and\ \citenamefont {Godfrey}}]{Harris}%
  \BibitemOpen
  \bibfield  {author} {\bibinfo {author} {\bibfnamefont {M.~J.}\ \bibnamefont
  {Harris}}, \bibinfo {author} {\bibfnamefont {S.~T.}\ \bibnamefont
  {Bramwell}}, \bibinfo {author} {\bibfnamefont {D.~F.}\ \bibnamefont
  {McMorrow}}, \bibinfo {author} {\bibfnamefont {T.}~\bibnamefont {Zeiske}}, \
  and\ \bibinfo {author} {\bibfnamefont {K.~W.}\ \bibnamefont {Godfrey}},\
  }\bibfield  {title} {\enquote {\bibinfo {title} {Geometrical frustration in
  the ferromagnetic pyrochlore $\textup{Ho}_2\textup{Ti}_2\textup{O}_7$},}\
  }\href {\doibase 10.1103/PhysRevLett.79.2554} {\bibfield  {journal} {\bibinfo
   {journal} {Phys. Rev. Lett.}\ }\textbf {\bibinfo {volume} {79}},\ \bibinfo
  {pages} {2554--2557} (\bibinfo {year} {1997})}\BibitemShut {NoStop}%
\bibitem [{\citenamefont {Bramwell}\ and\ \citenamefont
  {Harris}(1998)}]{BramwellHarris}%
  \BibitemOpen
  \bibfield  {author} {\bibinfo {author} {\bibfnamefont {S.~T.}\ \bibnamefont
  {Bramwell}}\ and\ \bibinfo {author} {\bibfnamefont {M.~J.}\ \bibnamefont
  {Harris}},\ }\bibfield  {title} {\enquote {\bibinfo {title} {Frustration in
  $\textup{I}$sing-type spin models on the pyrochlore lattice},}\ }\href
  {\doibase 10.1088/0953-8984/10/14/002} {\bibfield  {journal} {\bibinfo
  {journal} {Journal of Physics: Condensed Matter}\ }\textbf {\bibinfo {volume}
  {10}},\ \bibinfo {pages} {L215--L220} (\bibinfo {year} {1998})}\BibitemShut
  {NoStop}%
\bibitem [{\citenamefont {den Hertog}\ and\ \citenamefont
  {Gingras}(2000)}]{DenHertog}%
  \BibitemOpen
  \bibfield  {author} {\bibinfo {author} {\bibfnamefont {B.~C.}\ \bibnamefont
  {den Hertog}}\ and\ \bibinfo {author} {\bibfnamefont {M.~J.~P.}\ \bibnamefont
  {Gingras}},\ }\bibfield  {title} {\enquote {\bibinfo {title} {Dipolar
  interactions and origin of spin ice in $\textup{I}$sing pyrochlore
  magnets},}\ }\href {\doibase 10.1103/PhysRevLett.84.3430} {\bibfield
  {journal} {\bibinfo  {journal} {Phys. Rev. Lett.}\ }\textbf {\bibinfo
  {volume} {84}},\ \bibinfo {pages} {3430--3433} (\bibinfo {year}
  {2000})}\BibitemShut {NoStop}%
\bibitem [{\citenamefont {Yavors'kii}\ \emph {et~al.}(2008)\citenamefont
  {Yavors'kii}, \citenamefont {Fennell}, \citenamefont {Gingras},\ and\
  \citenamefont {Bramwell}}]{Yavorskii}%
  \BibitemOpen
  \bibfield  {author} {\bibinfo {author} {\bibfnamefont {T.}~\bibnamefont
  {Yavors'kii}}, \bibinfo {author} {\bibfnamefont {T.}~\bibnamefont {Fennell}},
  \bibinfo {author} {\bibfnamefont {M.~J.~P.}\ \bibnamefont {Gingras}}, \ and\
  \bibinfo {author} {\bibfnamefont {S.~T.}\ \bibnamefont {Bramwell}},\
  }\bibfield  {title} {\enquote {\bibinfo {title}
  {$\textup{Dy}_2\textup{Ti}_2\textup{O}_7$ spin ice: A test case for emergent
  clusters in a frustrated magnet},}\ }\href {\doibase
  10.1103/PhysRevLett.101.037204} {\bibfield  {journal} {\bibinfo  {journal}
  {Phys. Rev. Lett.}\ }\textbf {\bibinfo {volume} {101}},\ \bibinfo {pages}
  {037204} (\bibinfo {year} {2008})}\BibitemShut {NoStop}%
\bibitem [{\citenamefont {Ryzhkin}(2005)}]{Ryzhkin}%
  \BibitemOpen
  \bibfield  {author} {\bibinfo {author} {\bibfnamefont {I.~A.}\ \bibnamefont
  {Ryzhkin}},\ }\bibfield  {title} {\enquote {\bibinfo {title} {Magnetic
  relaxation in rare-earth oxide pyrochlores},}\ }\href {\doibase
  10.1134/1.2103216} {\bibfield  {journal} {\bibinfo  {journal} {Journal of
  Experimental and Theoretical Physics}\ }\textbf {\bibinfo {volume} {101}},\
  \bibinfo {pages} {481--486} (\bibinfo {year} {2005})}\BibitemShut {NoStop}%
\bibitem [{\citenamefont {Castelnovo}\ \emph {et~al.}(2008)\citenamefont
  {Castelnovo}, \citenamefont {Moessner},\ and\ \citenamefont {Sondhi}}]{CMS}%
  \BibitemOpen
  \bibfield  {author} {\bibinfo {author} {\bibfnamefont {C.}~\bibnamefont
  {Castelnovo}}, \bibinfo {author} {\bibfnamefont {R.}~\bibnamefont
  {Moessner}}, \ and\ \bibinfo {author} {\bibfnamefont {S.~L.}\ \bibnamefont
  {Sondhi}},\ }\bibfield  {title} {\enquote {\bibinfo {title} {Magnetic
  monopoles in spin ice},}\ }\href {\doibase 10.1038/nature06433} {\bibfield
  {journal} {\bibinfo  {journal} {Nature}\ }\textbf {\bibinfo {volume} {451}},\
  \bibinfo {pages} {42--45} (\bibinfo {year} {2008})}\BibitemShut {NoStop}%
\bibitem [{\citenamefont {Brooks-Bartlett}\ \emph {et~al.}(2014)\citenamefont
  {Brooks-Bartlett}, \citenamefont {Banks}, \citenamefont {Jaubert},
  \citenamefont {Harman-Clarke},\ and\ \citenamefont
  {Holdsworth}}]{BrooksBartlett}%
  \BibitemOpen
  \bibfield  {author} {\bibinfo {author} {\bibfnamefont {M.~E.}\ \bibnamefont
  {Brooks-Bartlett}}, \bibinfo {author} {\bibfnamefont {S.~T.}\ \bibnamefont
  {Banks}}, \bibinfo {author} {\bibfnamefont {L.~D.~C.}\ \bibnamefont
  {Jaubert}}, \bibinfo {author} {\bibfnamefont {A.}~\bibnamefont
  {Harman-Clarke}}, \ and\ \bibinfo {author} {\bibfnamefont {P.~C.~W.}\
  \bibnamefont {Holdsworth}},\ }\bibfield  {title} {\enquote {\bibinfo {title}
  {Magnetic-moment fragmentation and monopole crystallization},}\ }\href
  {\doibase 10.1103/PhysRevX.4.011007} {\bibfield  {journal} {\bibinfo
  {journal} {Phys. Rev. X}\ }\textbf {\bibinfo {volume} {4}},\ \bibinfo {pages}
  {011007} (\bibinfo {year} {2014})}\BibitemShut {NoStop}%
\bibitem [{\citenamefont {Petit}\ \emph {et~al.}(2016)\citenamefont {Petit},
  \citenamefont {Lhotel}, \citenamefont {Canals}, \citenamefont
  {Ciomaga~Hatnean}, \citenamefont {Ollivier}, \citenamefont {Mutka},
  \citenamefont {Ressouche}, \citenamefont {Wildes}, \citenamefont {Lees},\
  and\ \citenamefont {Balakrishnan}}]{Petit}%
  \BibitemOpen
  \bibfield  {author} {\bibinfo {author} {\bibfnamefont {S.}~\bibnamefont
  {Petit}}, \bibinfo {author} {\bibfnamefont {E.}~\bibnamefont {Lhotel}},
  \bibinfo {author} {\bibfnamefont {B.}~\bibnamefont {Canals}}, \bibinfo
  {author} {\bibfnamefont {M.}~\bibnamefont {Ciomaga~Hatnean}}, \bibinfo
  {author} {\bibfnamefont {J.}~\bibnamefont {Ollivier}}, \bibinfo {author}
  {\bibfnamefont {H.}~\bibnamefont {Mutka}}, \bibinfo {author} {\bibfnamefont
  {E.}~\bibnamefont {Ressouche}}, \bibinfo {author} {\bibfnamefont {A.~R.}\
  \bibnamefont {Wildes}}, \bibinfo {author} {\bibfnamefont {M.~R.}\
  \bibnamefont {Lees}}, \ and\ \bibinfo {author} {\bibfnamefont
  {G.}~\bibnamefont {Balakrishnan}},\ }\bibfield  {title} {\enquote {\bibinfo
  {title} {Observation of magnetic fragmentation in spin ice},}\ }\href
  {https://doi.org/10.1038/nphys3710} {\bibfield  {journal} {\bibinfo
  {journal} {Nature Physics}\ }\textbf {\bibinfo {volume} {12}},\ \bibinfo
  {pages} {746} (\bibinfo {year} {2016})}\BibitemShut {NoStop}%
\bibitem [{\citenamefont {Castelnovo}\ \emph {et~al.}(2011)\citenamefont
  {Castelnovo}, \citenamefont {Moessner},\ and\ \citenamefont
  {Sondhi}}]{CMSDH}%
  \BibitemOpen
  \bibfield  {author} {\bibinfo {author} {\bibfnamefont {C.}~\bibnamefont
  {Castelnovo}}, \bibinfo {author} {\bibfnamefont {R.}~\bibnamefont
  {Moessner}}, \ and\ \bibinfo {author} {\bibfnamefont {S.~L.}\ \bibnamefont
  {Sondhi}},\ }\bibfield  {title} {\enquote {\bibinfo {title}
  {Debye-$\textup{H}$\"uckel theory for spin ice at low temperature},}\ }\href
  {\doibase 10.1103/PhysRevB.84.144435} {\bibfield  {journal} {\bibinfo
  {journal} {Phys. Rev. B}\ }\textbf {\bibinfo {volume} {84}},\ \bibinfo
  {pages} {144435} (\bibinfo {year} {2011})}\BibitemShut {NoStop}%
\bibitem [{\citenamefont {Kaiser}\ \emph {et~al.}(2018)\citenamefont {Kaiser},
  \citenamefont {Bloxsom}, \citenamefont {Bovo}, \citenamefont {Bramwell},
  \citenamefont {Holdsworth},\ and\ \citenamefont {Moessner}}]{Kaiser}%
  \BibitemOpen
  \bibfield  {author} {\bibinfo {author} {\bibfnamefont {V.}~\bibnamefont
  {Kaiser}}, \bibinfo {author} {\bibfnamefont {J.}~\bibnamefont {Bloxsom}},
  \bibinfo {author} {\bibfnamefont {L.}~\bibnamefont {Bovo}}, \bibinfo {author}
  {\bibfnamefont {S.~T.}\ \bibnamefont {Bramwell}}, \bibinfo {author}
  {\bibfnamefont {P.~C.~W.}\ \bibnamefont {Holdsworth}}, \ and\ \bibinfo
  {author} {\bibfnamefont {R.}~\bibnamefont {Moessner}},\ }\bibfield  {title}
  {\enquote {\bibinfo {title} {Emergent electrochemistry in spin ice:
  Debye-$\textup{H}$\"uckel theory and beyond},}\ }\href {\doibase
  10.1103/PhysRevB.98.144413} {\bibfield  {journal} {\bibinfo  {journal} {Phys.
  Rev. B}\ }\textbf {\bibinfo {volume} {98}},\ \bibinfo {pages} {144413}
  (\bibinfo {year} {2018})}\BibitemShut {NoStop}%
\bibitem [{\citenamefont {Kaiser}\ \emph {et~al.}(2015)\citenamefont {Kaiser},
  \citenamefont {Bramwell}, \citenamefont {Holdsworth},\ and\ \citenamefont
  {Moessner}}]{KaiserPRL}%
  \BibitemOpen
  \bibfield  {author} {\bibinfo {author} {\bibfnamefont {V.}~\bibnamefont
  {Kaiser}}, \bibinfo {author} {\bibfnamefont {S.~T.}\ \bibnamefont
  {Bramwell}}, \bibinfo {author} {\bibfnamefont {P.~C.~W.}\ \bibnamefont
  {Holdsworth}}, \ and\ \bibinfo {author} {\bibfnamefont {R.}~\bibnamefont
  {Moessner}},\ }\bibfield  {title} {\enquote {\bibinfo {title} {ac
  $\textup{W}$ien effect in spin ice, manifest in nonlinear, nonequilibrium
  susceptibility},}\ }\href {\doibase 10.1103/PhysRevLett.115.037201}
  {\bibfield  {journal} {\bibinfo  {journal} {Phys. Rev. Lett.}\ }\textbf
  {\bibinfo {volume} {115}},\ \bibinfo {pages} {037201} (\bibinfo {year}
  {2015})}\BibitemShut {NoStop}%
\bibitem [{\citenamefont {Bovo}\ \emph
  {et~al.}(2013{\natexlab{a}})\citenamefont {Bovo}, \citenamefont {Bloxsom},
  \citenamefont {Prabhakaran}, \citenamefont {Aeppli},\ and\ \citenamefont
  {Bramwell}}]{Bovo}%
  \BibitemOpen
  \bibfield  {author} {\bibinfo {author} {\bibfnamefont {L.}~\bibnamefont
  {Bovo}}, \bibinfo {author} {\bibfnamefont {J.~A.}\ \bibnamefont {Bloxsom}},
  \bibinfo {author} {\bibfnamefont {D.}~\bibnamefont {Prabhakaran}}, \bibinfo
  {author} {\bibfnamefont {G.}~\bibnamefont {Aeppli}}, \ and\ \bibinfo {author}
  {\bibfnamefont {S.~T.}\ \bibnamefont {Bramwell}},\ }\bibfield  {title}
  {\enquote {\bibinfo {title} {Brownian motion and quantum dynamics of magnetic
  monopoles in spin ice},}\ }\href {https://doi.org/10.1038/ncomms2551}
  {\bibfield  {journal} {\bibinfo  {journal} {Nature Communications}\ }\textbf
  {\bibinfo {volume} {4}},\ \bibinfo {pages} {1535} (\bibinfo {year}
  {2013}{\natexlab{a}})},\ \bibinfo {note} {article}\BibitemShut {NoStop}%
\bibitem [{\citenamefont {Paulsen}\ \emph {et~al.}(2016)\citenamefont
  {Paulsen}, \citenamefont {Giblin}, \citenamefont {Lhotel}, \citenamefont
  {Prabhakaran}, \citenamefont {Balakrishnan}, \citenamefont {Matsuhira},\ and\
  \citenamefont {Bramwell}}]{Paulsen}%
  \BibitemOpen
  \bibfield  {author} {\bibinfo {author} {\bibfnamefont {C.}~\bibnamefont
  {Paulsen}}, \bibinfo {author} {\bibfnamefont {S.~R.}\ \bibnamefont {Giblin}},
  \bibinfo {author} {\bibfnamefont {E.}~\bibnamefont {Lhotel}}, \bibinfo
  {author} {\bibfnamefont {D.}~\bibnamefont {Prabhakaran}}, \bibinfo {author}
  {\bibfnamefont {G.}~\bibnamefont {Balakrishnan}}, \bibinfo {author}
  {\bibfnamefont {K.}~\bibnamefont {Matsuhira}}, \ and\ \bibinfo {author}
  {\bibfnamefont {S.~T.}\ \bibnamefont {Bramwell}},\ }\bibfield  {title}
  {\enquote {\bibinfo {title} {Experimental signature of the attractive
  $\textup{C}$oulomb force between positive and negative magnetic monopoles in
  spin ice},}\ }\href {https://doi.org/10.1038/nphys3704} {\bibfield  {journal}
  {\bibinfo  {journal} {Nature Physics}\ }\textbf {\bibinfo {volume} {12}},\
  \bibinfo {pages} {661} (\bibinfo {year} {2016})}\BibitemShut {NoStop}%
\bibitem [{\citenamefont {Lee}\ \emph {et~al.}(2015)\citenamefont {Lee},
  \citenamefont {Vella}, \citenamefont {Perkin},\ and\ \citenamefont
  {Goriely}}]{Lee2015}%
  \BibitemOpen
  \bibfield  {author} {\bibinfo {author} {\bibfnamefont {A.~A.}\ \bibnamefont
  {Lee}}, \bibinfo {author} {\bibfnamefont {D.}~\bibnamefont {Vella}}, \bibinfo
  {author} {\bibfnamefont {S.}~\bibnamefont {Perkin}}, \ and\ \bibinfo {author}
  {\bibfnamefont {A.}~\bibnamefont {Goriely}},\ }\bibfield  {title} {\enquote
  {\bibinfo {title} {Are room-temperature ionic liquids dilute electrolytes?}}\
  }\href {\doibase 10.1021/jz502250z} {\bibfield  {journal} {\bibinfo
  {journal} {The Journal of Physical Chemistry Letters}\ }\textbf {\bibinfo
  {volume} {6}},\ \bibinfo {pages} {159--163} (\bibinfo {year}
  {2015})}\BibitemShut {NoStop}%
\bibitem [{\citenamefont {Lee}\ \emph {et~al.}(2017)\citenamefont {Lee},
  \citenamefont {Perez-Martinez}, \citenamefont {Smith},\ and\ \citenamefont
  {Perkin}}]{Lee}%
  \BibitemOpen
  \bibfield  {author} {\bibinfo {author} {\bibfnamefont {A.~A.}\ \bibnamefont
  {Lee}}, \bibinfo {author} {\bibfnamefont {C.~S.}\ \bibnamefont
  {Perez-Martinez}}, \bibinfo {author} {\bibfnamefont {A.~M.}\ \bibnamefont
  {Smith}}, \ and\ \bibinfo {author} {\bibfnamefont {S.}~\bibnamefont
  {Perkin}},\ }\bibfield  {title} {\enquote {\bibinfo {title} {Underscreening
  in concentrated electrolytes},}\ }\href {\doibase 10.1039/C6FD00250A}
  {\bibfield  {journal} {\bibinfo  {journal} {Faraday Discuss.}\ }\textbf
  {\bibinfo {volume} {199}},\ \bibinfo {pages} {239--259} (\bibinfo {year}
  {2017})}\BibitemShut {NoStop}%
\bibitem [{\citenamefont {Marshall}\ and\ \citenamefont
  {Lowde}(1968)}]{Marshall-Lowde}%
  \BibitemOpen
  \bibfield  {author} {\bibinfo {author} {\bibfnamefont {W.}~\bibnamefont
  {Marshall}}\ and\ \bibinfo {author} {\bibfnamefont {R.~D.}\ \bibnamefont
  {Lowde}},\ }\bibfield  {title} {\enquote {\bibinfo {title} {Magnetic
  correlations and neutron scattering},}\ }\href {\doibase
  10.1088/0034-4885/31/2/305} {\bibfield  {journal} {\bibinfo  {journal}
  {Reports on Progress in Physics}\ }\textbf {\bibinfo {volume} {31}},\
  \bibinfo {pages} {705--775} (\bibinfo {year} {1968})}\BibitemShut {NoStop}%
\bibitem [{\citenamefont {Bramwell}(2017)}]{Bramwell_NComms}%
  \BibitemOpen
  \bibfield  {author} {\bibinfo {author} {\bibfnamefont {S.~T.}\ \bibnamefont
  {Bramwell}},\ }\bibfield  {title} {\enquote {\bibinfo {title} {Harmonic phase
  in polar liquids and spin ice},}\ }\href {\doibase
  10.1038/s41467-017-02102-1} {\bibfield  {journal} {\bibinfo  {journal}
  {Nature Communications}\ }\textbf {\bibinfo {volume} {8}},\ \bibinfo {pages}
  {2088} (\bibinfo {year} {2017})}\BibitemShut {NoStop}%
\bibitem [{\citenamefont {Nye}(1985)}]{Nye}%
  \BibitemOpen
  \bibfield  {author} {\bibinfo {author} {\bibfnamefont {J.~F.}\ \bibnamefont
  {Nye}},\ }\href {https://books.google.se/books?id=ugwql-uVB44C} {\emph
  {\bibinfo {title} {Physical Properties of Crystals}}},\ Oxford science
  publications\ (\bibinfo  {publisher} {Clarendon Press},\ \bibinfo {year}
  {1985})\BibitemShut {NoStop}%
\bibitem [{\citenamefont {Onsager}(1936)}]{Onsager}%
  \BibitemOpen
  \bibfield  {author} {\bibinfo {author} {\bibfnamefont {L.}~\bibnamefont
  {Onsager}},\ }\bibfield  {title} {\enquote {\bibinfo {title} {Electric
  moments of molecules in liquids},}\ }\href {\doibase 10.1021/ja01299a050}
  {\bibfield  {journal} {\bibinfo  {journal} {Journal of the American Chemical
  Society}\ }\textbf {\bibinfo {volume} {58}},\ \bibinfo {pages} {1486--1493}
  (\bibinfo {year} {1936})}\BibitemShut {NoStop}%
\bibitem [{\citenamefont {Twengstr{\"o}m}\ \emph {et~al.}(2019)\citenamefont
  {Twengstr{\"o}m}, \citenamefont {Bovo}, \citenamefont {Petrenko},
  \citenamefont {Ouladdiaf}, \citenamefont {Fennell}, \citenamefont
  {Henelius},\ and\ \citenamefont {Bramwell}}]{ZoneCenterPaper}%
  \BibitemOpen
  \bibfield  {author} {\bibinfo {author} {\bibfnamefont {M.}~\bibnamefont
  {Twengstr{\"o}m}}, \bibinfo {author} {\bibfnamefont {L.}~\bibnamefont
  {Bovo}}, \bibinfo {author} {\bibfnamefont {O.~A.}\ \bibnamefont {Petrenko}},
  \bibinfo {author} {\bibfnamefont {B.}~\bibnamefont {Ouladdiaf}}, \bibinfo
  {author} {\bibfnamefont {T.}~\bibnamefont {Fennell}}, \bibinfo {author}
  {\bibfnamefont {P.}~\bibnamefont {Henelius}}, \ and\ \bibinfo {author}
  {\bibfnamefont {S.~T.}\ \bibnamefont {Bramwell}},\ }\bibfield  {title}
  {\enquote {\bibinfo {title} {{Zone center physics in magnetic diffuse neutron
  scattering}},}\ }\href@noop {} {\bibfield  {journal} {\bibinfo  {journal} {To
  be submitted}\ } (\bibinfo {year} {2019})}\BibitemShut {NoStop}%
\bibitem [{\citenamefont {Subramanian}\ \emph {et~al.}(1983)\citenamefont
  {Subramanian}, \citenamefont {Aravamudan},\ and\ \citenamefont
  {Subba~Rao}}]{Pyrochlore}%
  \BibitemOpen
  \bibfield  {author} {\bibinfo {author} {\bibfnamefont {M.~A.}\ \bibnamefont
  {Subramanian}}, \bibinfo {author} {\bibfnamefont {G.}~\bibnamefont
  {Aravamudan}}, \ and\ \bibinfo {author} {\bibfnamefont {G.~V.}\ \bibnamefont
  {Subba~Rao}},\ }\bibfield  {title} {\enquote {\bibinfo {title} {{Oxide
  pyrochlores — A review}},}\ }\href@noop {} {\bibfield  {journal} {\bibinfo
  {journal} {Progress in Solid State Chemistry}\ }\textbf {\bibinfo {volume}
  {15}},\ \bibinfo {pages} {55 -- 143} (\bibinfo {year} {1983})}\BibitemShut
  {NoStop}%
\bibitem [{\citenamefont {Rosenkranz}\ \emph {et~al.}(2000)\citenamefont
  {Rosenkranz}, \citenamefont {Ramirez}, \citenamefont {Hayashi}, \citenamefont
  {Cava}, \citenamefont {Siddharthan},\ and\ \citenamefont
  {Shastry}}]{crystal1}%
  \BibitemOpen
  \bibfield  {author} {\bibinfo {author} {\bibfnamefont {S.}~\bibnamefont
  {Rosenkranz}}, \bibinfo {author} {\bibfnamefont {A.~P.}\ \bibnamefont
  {Ramirez}}, \bibinfo {author} {\bibfnamefont {A.}~\bibnamefont {Hayashi}},
  \bibinfo {author} {\bibfnamefont {R.~J.}\ \bibnamefont {Cava}}, \bibinfo
  {author} {\bibfnamefont {R.}~\bibnamefont {Siddharthan}}, \ and\ \bibinfo
  {author} {\bibfnamefont {B.~S.}\ \bibnamefont {Shastry}},\ }\bibfield
  {title} {\enquote {\bibinfo {title} {Crystal-field interaction in the
  pyrochlore magnet $\textup{Ho}_2\textup{Ti}_2\textup{O}_7$},}\ }\href@noop {}
  {\bibfield  {journal} {\bibinfo  {journal} {Journal of Applied Physics}\
  }\textbf {\bibinfo {volume} {87}},\ \bibinfo {pages} {5914--5916} (\bibinfo
  {year} {2000})}\BibitemShut {NoStop}%
\bibitem [{\citenamefont {Bertin}\ \emph {et~al.}(2012)\citenamefont {Bertin},
  \citenamefont {Chapuis}, \citenamefont {Dalmas~de Réotier},\ and\
  \citenamefont {Yaouanc}}]{crystal2}%
  \BibitemOpen
  \bibfield  {author} {\bibinfo {author} {\bibfnamefont {A.}~\bibnamefont
  {Bertin}}, \bibinfo {author} {\bibfnamefont {Y.}~\bibnamefont {Chapuis}},
  \bibinfo {author} {\bibfnamefont {P.}~\bibnamefont {Dalmas~de Réotier}}, \
  and\ \bibinfo {author} {\bibfnamefont {A.}~\bibnamefont {Yaouanc}},\
  }\bibfield  {title} {\enquote {\bibinfo {title} {Crystal electric field in
  the $\textup{R}_2\textup{Ti}_2\textup{O}_7$ pyrochlore compounds},}\
  }\href@noop {} {\bibfield  {journal} {\bibinfo  {journal} {Journal of
  Physics: Condensed Matter}\ }\textbf {\bibinfo {volume} {24}},\ \bibinfo
  {pages} {256003} (\bibinfo {year} {2012})}\BibitemShut {NoStop}%
\bibitem [{\citenamefont {Jana}\ \emph {et~al.}(2002)\citenamefont {Jana},
  \citenamefont {Sengupta},\ and\ \citenamefont {Ghosh}}]{crystal3}%
  \BibitemOpen
  \bibfield  {author} {\bibinfo {author} {\bibfnamefont {Y.~M.}\ \bibnamefont
  {Jana}}, \bibinfo {author} {\bibfnamefont {A.}~\bibnamefont {Sengupta}}, \
  and\ \bibinfo {author} {\bibfnamefont {D.}~\bibnamefont {Ghosh}},\ }\bibfield
   {title} {\enquote {\bibinfo {title} {{Estimation of single ion anisotropy in
  pyrochlore $\textup{Dy}_2\textup{Ti}_2\textup{O}_7$, a geometrically
  frustrated system, using crystal field theory}},}\ }\href@noop {} {\bibfield
  {journal} {\bibinfo  {journal} {Journal of Magnetism and Magnetic Materials}\
  }\textbf {\bibinfo {volume} {248}},\ \bibinfo {pages} {7 -- 18} (\bibinfo
  {year} {2002})}\BibitemShut {NoStop}%
\bibitem [{\citenamefont {Rau}\ and\ \citenamefont
  {Gingras}(2015{\natexlab{a}})}]{Rau_Gingras_2015}%
  \BibitemOpen
  \bibfield  {author} {\bibinfo {author} {\bibfnamefont {J.~G.}\ \bibnamefont
  {Rau}}\ and\ \bibinfo {author} {\bibfnamefont {M.~J.~P.}\ \bibnamefont
  {Gingras}},\ }\bibfield  {title} {\enquote {\bibinfo {title} {{Magnitude of
  quantum effects in classical spin ices}},}\ }\href@noop {} {\bibfield
  {journal} {\bibinfo  {journal} {Phys. Rev. B}\ }\textbf {\bibinfo {volume}
  {92}},\ \bibinfo {pages} {144417} (\bibinfo {year}
  {2015}{\natexlab{a}})}\BibitemShut {NoStop}%
\bibitem [{\citenamefont {Melko}\ and\ \citenamefont
  {Gingras}(2004)}]{melko04}%
  \BibitemOpen
  \bibfield  {author} {\bibinfo {author} {\bibfnamefont {R.~G.}\ \bibnamefont
  {Melko}}\ and\ \bibinfo {author} {\bibfnamefont {M.~J.~P.}\ \bibnamefont
  {Gingras}},\ }\bibfield  {title} {\enquote {\bibinfo {title} {{Monte Carlo
  studies of the dipolar spin ice model}},}\ }\href@noop {} {\bibfield
  {journal} {\bibinfo  {journal} {J. Phys.: Condens. Matter}\ }\textbf
  {\bibinfo {volume} {16}},\ \bibinfo {pages} {R1277} (\bibinfo {year}
  {2004})}\BibitemShut {NoStop}%
\bibitem [{\citenamefont {Twengstr\"om}\ \emph {et~al.}(2017)\citenamefont
  {Twengstr\"om}, \citenamefont {Bovo}, \citenamefont {Gingras}, \citenamefont
  {Bramwell},\ and\ \citenamefont {Henelius}}]{prm_twengstrom_2017}%
  \BibitemOpen
  \bibfield  {author} {\bibinfo {author} {\bibfnamefont {M.}~\bibnamefont
  {Twengstr\"om}}, \bibinfo {author} {\bibfnamefont {L.}~\bibnamefont {Bovo}},
  \bibinfo {author} {\bibfnamefont {M.~J.~P.}\ \bibnamefont {Gingras}},
  \bibinfo {author} {\bibfnamefont {S.~T.}\ \bibnamefont {Bramwell}}, \ and\
  \bibinfo {author} {\bibfnamefont {P.}~\bibnamefont {Henelius}},\ }\bibfield
  {title} {\enquote {\bibinfo {title} {Microscopic aspects of magnetic lattice
  demagnetizing factors},}\ }\href@noop {} {\bibfield  {journal} {\bibinfo
  {journal} {Phys. Rev. Materials}\ }\textbf {\bibinfo {volume} {1}},\ \bibinfo
  {pages} {044406} (\bibinfo {year} {2017})}\BibitemShut {NoStop}%
\bibitem [{\citenamefont {Ewald}(1921)}]{ewald21}%
  \BibitemOpen
  \bibfield  {author} {\bibinfo {author} {\bibfnamefont {P.~P.}\ \bibnamefont
  {Ewald}},\ }\bibfield  {title} {\enquote {\bibinfo {title} {{Die Berechnung
  optischer und elektrostatischer Gitterpotentiale}},}\ }\href@noop {}
  {\bibfield  {journal} {\bibinfo  {journal} {Annalen der Physik}\ }\textbf
  {\bibinfo {volume} {369}},\ \bibinfo {pages} {253--287} (\bibinfo {year}
  {1921})}\BibitemShut {NoStop}%
\bibitem [{\citenamefont {Melko}\ \emph {et~al.}(2001)\citenamefont {Melko},
  \citenamefont {den Hertog},\ and\ \citenamefont {Gingras}}]{Melko}%
  \BibitemOpen
  \bibfield  {author} {\bibinfo {author} {\bibfnamefont {R.~G.}\ \bibnamefont
  {Melko}}, \bibinfo {author} {\bibfnamefont {B.~C.}\ \bibnamefont {den
  Hertog}}, \ and\ \bibinfo {author} {\bibfnamefont {M.~J.~P.}\ \bibnamefont
  {Gingras}},\ }\bibfield  {title} {\enquote {\bibinfo {title} {Long-range
  order at low temperatures in dipolar spin ice},}\ }\href {\doibase
  10.1103/PhysRevLett.87.067203} {\bibfield  {journal} {\bibinfo  {journal}
  {Phys. Rev. Lett.}\ }\textbf {\bibinfo {volume} {87}},\ \bibinfo {pages}
  {067203} (\bibinfo {year} {2001})}\BibitemShut {NoStop}%
\bibitem [{\citenamefont {Twengstr{\"o}m}(2018)}]{mtthesis}%
  \BibitemOpen
  \bibfield  {author} {\bibinfo {author} {\bibfnamefont {M.}~\bibnamefont
  {Twengstr{\"o}m}},\ }\emph {\bibinfo {title} {Spin ice and demagnetising
  theory}},\ \href@noop {} {Ph.D. thesis},\ \bibinfo  {school} {KTH Royal
  Institute of Technology} (\bibinfo {year} {2018})\BibitemShut {NoStop}%
\bibitem [{\citenamefont {Bramwell}(2012)}]{Bramwell_PhilTrans}%
  \BibitemOpen
  \bibfield  {author} {\bibinfo {author} {\bibfnamefont {S.~T.}\ \bibnamefont
  {Bramwell}},\ }\bibfield  {title} {\enquote {\bibinfo {title} {Generalized
  longitudinal susceptibility for magnetic monopoles in spin ice},}\ }\href
  {\doibase 10.1098/rsta.2011.0596} {\bibfield  {journal} {\bibinfo  {journal}
  {Philosophical Transactions of the Royal Society A: Mathematical, Physical
  and Engineering Sciences}\ }\textbf {\bibinfo {volume} {370}},\ \bibinfo
  {pages} {5738--5766} (\bibinfo {year} {2012})}\BibitemShut {NoStop}%
\bibitem [{\citenamefont {Jaubert}\ \emph {et~al.}(2013)\citenamefont
  {Jaubert}, \citenamefont {Harris}, \citenamefont {Fennell}, \citenamefont
  {Melko}, \citenamefont {Bramwell},\ and\ \citenamefont
  {Holdsworth}}]{Jaubert}%
  \BibitemOpen
  \bibfield  {author} {\bibinfo {author} {\bibfnamefont {L.~D.~C.}\
  \bibnamefont {Jaubert}}, \bibinfo {author} {\bibfnamefont {M.~J.}\
  \bibnamefont {Harris}}, \bibinfo {author} {\bibfnamefont {T.}~\bibnamefont
  {Fennell}}, \bibinfo {author} {\bibfnamefont {R.~G.}\ \bibnamefont {Melko}},
  \bibinfo {author} {\bibfnamefont {S.~T.}\ \bibnamefont {Bramwell}}, \ and\
  \bibinfo {author} {\bibfnamefont {P.~C.~W.}\ \bibnamefont {Holdsworth}},\
  }\bibfield  {title} {\enquote {\bibinfo {title} {Topological-sector
  fluctuations and $\textup{C}$urie-law crossover in spin ice},}\ }\href
  {\doibase 10.1103/PhysRevX.3.011014} {\bibfield  {journal} {\bibinfo
  {journal} {Phys. Rev. X}\ }\textbf {\bibinfo {volume} {3}},\ \bibinfo {pages}
  {011014} (\bibinfo {year} {2013})}\BibitemShut {NoStop}%
\bibitem [{\citenamefont {Morrish}(1980)}]{Morrish}%
  \BibitemOpen
  \bibfield  {author} {\bibinfo {author} {\bibfnamefont {A.~H.}\ \bibnamefont
  {Morrish}},\ }\href {https://books.google.se/books?id=ZjUbAQAAIAAJ} {\emph
  {\bibinfo {title} {The Physical Principles of Magnetism}}},\ Wiley series on
  the science and technology of materials\ (\bibinfo  {publisher} {R. E.
  Krieger Publishing Company},\ \bibinfo {year} {1980})\BibitemShut {NoStop}%
\bibitem [{\citenamefont {Bovo}\ \emph
  {et~al.}(2013{\natexlab{b}})\citenamefont {Bovo}, \citenamefont {Jaubert},
  \citenamefont {Holdsworth},\ and\ \citenamefont {Bramwell}}]{Bovo_Curie}%
  \BibitemOpen
  \bibfield  {author} {\bibinfo {author} {\bibfnamefont {L.}~\bibnamefont
  {Bovo}}, \bibinfo {author} {\bibfnamefont {L.~D.~C.}\ \bibnamefont
  {Jaubert}}, \bibinfo {author} {\bibfnamefont {P.~C.~W.}\ \bibnamefont
  {Holdsworth}}, \ and\ \bibinfo {author} {\bibfnamefont {S.~T.}\ \bibnamefont
  {Bramwell}},\ }\bibfield  {title} {\enquote {\bibinfo {title} {Crystal
  shape-dependent magnetic susceptibility and $\textup{C}$urie law crossover in
  the spin ices $\textup{Dy}_2\textup{Ti}_2\textup{O}_7$ and
  $\textup{Ho}_2\textup{Ti}_2\textup{O}_7$},}\ }\href {\doibase
  10.1088/0953-8984/25/38/386002} {\bibfield  {journal} {\bibinfo  {journal}
  {Journal of Physics: Condensed Matter}\ }\textbf {\bibinfo {volume} {25}},\
  \bibinfo {pages} {386002} (\bibinfo {year} {2013}{\natexlab{b}})}\BibitemShut
  {NoStop}%
\bibitem [{\citenamefont {Griffiths}(1982)}]{Griffiths}%
  \BibitemOpen
  \bibfield  {author} {\bibinfo {author} {\bibfnamefont {D.~J.}\ \bibnamefont
  {Griffiths}},\ }\bibfield  {title} {\enquote {\bibinfo {title} {Hyperfine
  splitting in the ground state of hydrogen},}\ }\href {\doibase
  10.1119/1.12733} {\bibfield  {journal} {\bibinfo  {journal} {American Journal
  of Physics}\ }\textbf {\bibinfo {volume} {50}},\ \bibinfo {pages} {698--703}
  (\bibinfo {year} {1982})}\BibitemShut {NoStop}%
\bibitem [{\citenamefont {Ryzhkin}(1984)}]{Ryzhkin_DSI}%
  \BibitemOpen
  \bibfield  {author} {\bibinfo {author} {\bibfnamefont {I.~A.}\ \bibnamefont
  {Ryzhkin}},\ }\bibfield  {title} {\enquote {\bibinfo {title} {Frustration
  model of proton disorder in ice},}\ }\href {\doibase
  https://doi.org/10.1016/0038-1098(84)90716-6} {\bibfield  {journal} {\bibinfo
   {journal} {Solid State Communications}\ }\textbf {\bibinfo {volume} {52}},\
  \bibinfo {pages} {49 -- 52} (\bibinfo {year} {1984})}\BibitemShut {NoStop}%
\bibitem [{\citenamefont {Sala}\ \emph {et~al.}(2014)\citenamefont {Sala},
  \citenamefont {Gutmann}, \citenamefont {Prabhakaran}, \citenamefont
  {Pomaranski}, \citenamefont {Mitchelitis}, \citenamefont {Kycia},
  \citenamefont {Porter}, \citenamefont {Castelnovo},\ and\ \citenamefont
  {Goff}}]{Sala}%
  \BibitemOpen
  \bibfield  {author} {\bibinfo {author} {\bibfnamefont {G.}~\bibnamefont
  {Sala}}, \bibinfo {author} {\bibfnamefont {M.~J.}\ \bibnamefont {Gutmann}},
  \bibinfo {author} {\bibfnamefont {D.}~\bibnamefont {Prabhakaran}}, \bibinfo
  {author} {\bibfnamefont {D.}~\bibnamefont {Pomaranski}}, \bibinfo {author}
  {\bibfnamefont {C.}~\bibnamefont {Mitchelitis}}, \bibinfo {author}
  {\bibfnamefont {J.~B.}\ \bibnamefont {Kycia}}, \bibinfo {author}
  {\bibfnamefont {D.~G.}\ \bibnamefont {Porter}}, \bibinfo {author}
  {\bibfnamefont {C.}~\bibnamefont {Castelnovo}}, \ and\ \bibinfo {author}
  {\bibfnamefont {J.~P.}\ \bibnamefont {Goff}},\ }\bibfield  {title} {\enquote
  {\bibinfo {title} {{Vacancy defects and monopole dynamics in oxygen-deficient
  pyrochlores}},}\ }\href@noop {} {\bibfield  {journal} {\bibinfo  {journal}
  {Nature Materials}\ }\textbf {\bibinfo {volume} {13}},\ \bibinfo {pages}
  {488--493} (\bibinfo {year} {2014})}\BibitemShut {NoStop}%
\bibitem [{\citenamefont {Revell}\ \emph {et~al.}(2012)\citenamefont {Revell},
  \citenamefont {Yaraskavitch}, \citenamefont {Mason}, \citenamefont {Ross},
  \citenamefont {Noad}, \citenamefont {Dabkowska}, \citenamefont {Gaulin},
  \citenamefont {Henelius},\ and\ \citenamefont {Kycia}}]{Revell}%
  \BibitemOpen
  \bibfield  {author} {\bibinfo {author} {\bibfnamefont {H.~M.}\ \bibnamefont
  {Revell}}, \bibinfo {author} {\bibfnamefont {L.~R.}\ \bibnamefont
  {Yaraskavitch}}, \bibinfo {author} {\bibfnamefont {J.~D.}\ \bibnamefont
  {Mason}}, \bibinfo {author} {\bibfnamefont {K.~A.}\ \bibnamefont {Ross}},
  \bibinfo {author} {\bibfnamefont {H.~M.~L.}\ \bibnamefont {Noad}}, \bibinfo
  {author} {\bibfnamefont {H.~A.}\ \bibnamefont {Dabkowska}}, \bibinfo {author}
  {\bibfnamefont {B.~D.}\ \bibnamefont {Gaulin}}, \bibinfo {author}
  {\bibfnamefont {P.}~\bibnamefont {Henelius}}, \ and\ \bibinfo {author}
  {\bibfnamefont {J.~B.}\ \bibnamefont {Kycia}},\ }\bibfield  {title} {\enquote
  {\bibinfo {title} {Evidence of impurity and boundary effects on magnetic
  monopole dynamics in spin ice},}\ }\href@noop {} {\bibfield  {journal}
  {\bibinfo  {journal} {Nature Physics}\ }\textbf {\bibinfo {volume} {9}},\
  \bibinfo {pages} {34} (\bibinfo {year} {2012})}\BibitemShut {NoStop}%
\bibitem [{\citenamefont {Ghasemi}\ \emph {et~al.}(2018)\citenamefont
  {Ghasemi}, \citenamefont {Scheie}, \citenamefont {Kindervater},\ and\
  \citenamefont {Koohpayeh}}]{Ghasemi}%
  \BibitemOpen
  \bibfield  {author} {\bibinfo {author} {\bibfnamefont {A.}~\bibnamefont
  {Ghasemi}}, \bibinfo {author} {\bibfnamefont {A.}~\bibnamefont {Scheie}},
  \bibinfo {author} {\bibfnamefont {J.}~\bibnamefont {Kindervater}}, \ and\
  \bibinfo {author} {\bibfnamefont {S.~M.}\ \bibnamefont {Koohpayeh}},\
  }\bibfield  {title} {\enquote {\bibinfo {title} {The pyrochlore
  $\textup{Ho}_2\textup{Ti}_2\textup{O}_7$: Synthesis, crystal growth, and
  stoichiometry},}\ }\href {\doibase
  https://doi.org/10.1016/j.jcrysgro.2018.08.006} {\bibfield  {journal}
  {\bibinfo  {journal} {Journal of Crystal Growth}\ }\textbf {\bibinfo {volume}
  {500}},\ \bibinfo {pages} {38 -- 43} (\bibinfo {year} {2018})}\BibitemShut
  {NoStop}%
\bibitem [{\citenamefont {Bovo}\ \emph {et~al.}(2018)\citenamefont {Bovo},
  \citenamefont {Twengstr{\"o}m}, \citenamefont {Petrenko}, \citenamefont
  {Fennell}, \citenamefont {Gingras}, \citenamefont {Bramwell},\ and\
  \citenamefont {Henelius}}]{Bovo_special}%
  \BibitemOpen
  \bibfield  {author} {\bibinfo {author} {\bibfnamefont {L.}~\bibnamefont
  {Bovo}}, \bibinfo {author} {\bibfnamefont {M.}~\bibnamefont
  {Twengstr{\"o}m}}, \bibinfo {author} {\bibfnamefont {O.~A.}\ \bibnamefont
  {Petrenko}}, \bibinfo {author} {\bibfnamefont {T.}~\bibnamefont {Fennell}},
  \bibinfo {author} {\bibfnamefont {M.~J.~P.}\ \bibnamefont {Gingras}},
  \bibinfo {author} {\bibfnamefont {S.~T.}\ \bibnamefont {Bramwell}}, \ and\
  \bibinfo {author} {\bibfnamefont {P.}~\bibnamefont {Henelius}},\ }\bibfield
  {title} {\enquote {\bibinfo {title} {Special temperatures in frustrated
  ferromagnets},}\ }\href {\doibase 10.1038/s41467-018-04297-3} {\bibfield
  {journal} {\bibinfo  {journal} {Nature Communications}\ }\textbf {\bibinfo
  {volume} {9}},\ \bibinfo {pages} {1999} (\bibinfo {year} {2018})}\BibitemShut
  {NoStop}%
\bibitem [{\citenamefont {Rau}\ and\ \citenamefont
  {Gingras}(2015{\natexlab{b}})}]{Gingras_quantum}%
  \BibitemOpen
  \bibfield  {author} {\bibinfo {author} {\bibfnamefont {J.~G.}\ \bibnamefont
  {Rau}}\ and\ \bibinfo {author} {\bibfnamefont {M.~J.~P.}\ \bibnamefont
  {Gingras}},\ }\bibfield  {title} {\enquote {\bibinfo {title} {Magnitude of
  quantum effects in classical spin ices},}\ }\href {\doibase
  10.1103/PhysRevB.92.144417} {\bibfield  {journal} {\bibinfo  {journal} {Phys.
  Rev. B}\ }\textbf {\bibinfo {volume} {92}},\ \bibinfo {pages} {144417}
  (\bibinfo {year} {2015}{\natexlab{b}})}\BibitemShut {NoStop}%
\bibitem [{\citenamefont {Paulsen}\ \emph {et~al.}(2019)\citenamefont
  {Paulsen}, \citenamefont {Giblin}, \citenamefont {Lhotel}, \citenamefont
  {Prabhakaran}, \citenamefont {Matsuhira}, \citenamefont {Balakrishnan},\ and\
  \citenamefont {Bramwell}}]{Paulsen_nuclear}%
  \BibitemOpen
  \bibfield  {author} {\bibinfo {author} {\bibfnamefont {C.}~\bibnamefont
  {Paulsen}}, \bibinfo {author} {\bibfnamefont {S.~R.}\ \bibnamefont {Giblin}},
  \bibinfo {author} {\bibfnamefont {E.}~\bibnamefont {Lhotel}}, \bibinfo
  {author} {\bibfnamefont {D.}~\bibnamefont {Prabhakaran}}, \bibinfo {author}
  {\bibfnamefont {K.}~\bibnamefont {Matsuhira}}, \bibinfo {author}
  {\bibfnamefont {G.}~\bibnamefont {Balakrishnan}}, \ and\ \bibinfo {author}
  {\bibfnamefont {S.~T.}\ \bibnamefont {Bramwell}},\ }\bibfield  {title}
  {\enquote {\bibinfo {title} {Nuclear spin assisted quantum tunnelling of
  magnetic monopoles in spin ice},}\ }\href {\doibase
  10.1038/s41467-019-09323-6} {\bibfield  {journal} {\bibinfo  {journal}
  {Nature Communications}\ }\textbf {\bibinfo {volume} {10}},\ \bibinfo {pages}
  {1509} (\bibinfo {year} {2019})}\BibitemShut {NoStop}%
\bibitem [{\citenamefont {Halpern}\ and\ \citenamefont
  {Holstein}(1941)}]{Halpern}%
  \BibitemOpen
  \bibfield  {author} {\bibinfo {author} {\bibfnamefont {O.}~\bibnamefont
  {Halpern}}\ and\ \bibinfo {author} {\bibfnamefont {T.}~\bibnamefont
  {Holstein}},\ }\bibfield  {title} {\enquote {\bibinfo {title} {On the passage
  of neutrons through ferromagnets},}\ }\href {\doibase 10.1103/PhysRev.59.960}
  {\bibfield  {journal} {\bibinfo  {journal} {Phys. Rev.}\ }\textbf {\bibinfo
  {volume} {59}},\ \bibinfo {pages} {960--981} (\bibinfo {year}
  {1941})}\BibitemShut {NoStop}%
\bibitem [{\citenamefont {Moon}\ \emph {et~al.}(1969)\citenamefont {Moon},
  \citenamefont {Riste},\ and\ \citenamefont {Koehler}}]{Moon}%
  \BibitemOpen
  \bibfield  {author} {\bibinfo {author} {\bibfnamefont {R.~M.}\ \bibnamefont
  {Moon}}, \bibinfo {author} {\bibfnamefont {T.}~\bibnamefont {Riste}}, \ and\
  \bibinfo {author} {\bibfnamefont {W.~C.}\ \bibnamefont {Koehler}},\
  }\bibfield  {title} {\enquote {\bibinfo {title} {Polarization analysis of
  thermal-neutron scattering},}\ }\href {\doibase 10.1103/PhysRev.181.920}
  {\bibfield  {journal} {\bibinfo  {journal} {Phys. Rev.}\ }\textbf {\bibinfo
  {volume} {181}},\ \bibinfo {pages} {920--931} (\bibinfo {year}
  {1969})}\BibitemShut {NoStop}%
\bibitem [{\citenamefont {Banks}\ and\ \citenamefont {Bramwell}(2012)}]{Banks}%
  \BibitemOpen
  \bibfield  {author} {\bibinfo {author} {\bibfnamefont {S.~T.}\ \bibnamefont
  {Banks}}\ and\ \bibinfo {author} {\bibfnamefont {S.~T.}\ \bibnamefont
  {Bramwell}},\ }\bibfield  {title} {\enquote {\bibinfo {title} {Magnetic
  frustration in the context of pseudo-dipolar ionic disorder},}\ }\href
  {\doibase 10.1209/0295-5075/97/27005} {\bibfield  {journal} {\bibinfo
  {journal} {{EPL} (Europhysics Letters)}\ }\textbf {\bibinfo {volume} {97}},\
  \bibinfo {pages} {27005} (\bibinfo {year} {2012})}\BibitemShut {NoStop}%
\bibitem [{\citenamefont {Benton}(2016)}]{Benton_frag}%
  \BibitemOpen
  \bibfield  {author} {\bibinfo {author} {\bibfnamefont {O.}~\bibnamefont
  {Benton}},\ }\bibfield  {title} {\enquote {\bibinfo {title} {Quantum origins
  of moment fragmentation in $\textup{Nd}_2\textup{Zr}_2\textup{O}_7$},}\
  }\href {\doibase 10.1103/PhysRevB.94.104430} {\bibfield  {journal} {\bibinfo
  {journal} {Phys. Rev. B}\ }\textbf {\bibinfo {volume} {94}},\ \bibinfo
  {pages} {104430} (\bibinfo {year} {2016})}\BibitemShut {NoStop}%
\bibitem [{\citenamefont {Yan}\ \emph {et~al.}(2018)\citenamefont {Yan},
  \citenamefont {Pohle},\ and\ \citenamefont {Shannon}}]{Yan}%
  \BibitemOpen
  \bibfield  {author} {\bibinfo {author} {\bibfnamefont {H.}~\bibnamefont
  {Yan}}, \bibinfo {author} {\bibfnamefont {R.}~\bibnamefont {Pohle}}, \ and\
  \bibinfo {author} {\bibfnamefont {N.}~\bibnamefont {Shannon}},\ }\bibfield
  {title} {\enquote {\bibinfo {title} {Half moons are pinch points with
  dispersion},}\ }\href {\doibase 10.1103/PhysRevB.98.140402} {\bibfield
  {journal} {\bibinfo  {journal} {Phys. Rev. B}\ }\textbf {\bibinfo {volume}
  {98}},\ \bibinfo {pages} {140402(R)} (\bibinfo {year} {2018})}\BibitemShut
  {NoStop}%
\bibitem [{\citenamefont {Benton}\ \emph
  {et~al.}(2016{\natexlab{a}})\citenamefont {Benton}, \citenamefont {Jaubert},
  \citenamefont {Yan},\ and\ \citenamefont {Shannon}}]{Benton_pinchline}%
  \BibitemOpen
  \bibfield  {author} {\bibinfo {author} {\bibfnamefont {O.}~\bibnamefont
  {Benton}}, \bibinfo {author} {\bibfnamefont {L.~D.~C.}\ \bibnamefont
  {Jaubert}}, \bibinfo {author} {\bibfnamefont {H.}~\bibnamefont {Yan}}, \ and\
  \bibinfo {author} {\bibfnamefont {N.}~\bibnamefont {Shannon}},\ }\bibfield
  {title} {\enquote {\bibinfo {title} {A spin-liquid with pinch-line
  singularities on the pyrochlore lattice},}\ }\href
  {https://doi.org/10.1038/ncomms11572} {\bibfield  {journal} {\bibinfo
  {journal} {Nature Communications}\ }\textbf {\bibinfo {volume} {7}},\
  \bibinfo {pages} {11572} (\bibinfo {year} {2016}{\natexlab{a}})},\ \bibinfo
  {note} {article}\BibitemShut {NoStop}%
\bibitem [{\citenamefont {Benton}\ \emph
  {et~al.}(2016{\natexlab{b}})\citenamefont {Benton}, \citenamefont {Sikora},\
  and\ \citenamefont {Shannon}}]{Benton_ice}%
  \BibitemOpen
  \bibfield  {author} {\bibinfo {author} {\bibfnamefont {O.}~\bibnamefont
  {Benton}}, \bibinfo {author} {\bibfnamefont {O.}~\bibnamefont {Sikora}}, \
  and\ \bibinfo {author} {\bibfnamefont {N.}~\bibnamefont {Shannon}},\
  }\bibfield  {title} {\enquote {\bibinfo {title} {Classical and quantum
  theories of proton disorder in hexagonal water ice},}\ }\href {\doibase
  10.1103/PhysRevB.93.125143} {\bibfield  {journal} {\bibinfo  {journal} {Phys.
  Rev. B}\ }\textbf {\bibinfo {volume} {93}},\ \bibinfo {pages} {125143}
  (\bibinfo {year} {2016}{\natexlab{b}})}\BibitemShut {NoStop}%
\bibitem [{\citenamefont {McClarty}\ \emph {et~al.}(2014)\citenamefont
  {McClarty}, \citenamefont {O'Brien},\ and\ \citenamefont
  {Pollmann}}]{McClarty}%
  \BibitemOpen
  \bibfield  {author} {\bibinfo {author} {\bibfnamefont {P.~A.}\ \bibnamefont
  {McClarty}}, \bibinfo {author} {\bibfnamefont {A.}~\bibnamefont {O'Brien}}, \
  and\ \bibinfo {author} {\bibfnamefont {F.}~\bibnamefont {Pollmann}},\
  }\bibfield  {title} {\enquote {\bibinfo {title} {$\textup{C}$oulombic charge
  ice},}\ }\href {\doibase 10.1103/PhysRevB.89.195123} {\bibfield  {journal}
  {\bibinfo  {journal} {Phys. Rev. B}\ }\textbf {\bibinfo {volume} {89}},\
  \bibinfo {pages} {195123} (\bibinfo {year} {2014})}\BibitemShut {NoStop}%
\bibitem [{\citenamefont {Slobinsky}\ \emph {et~al.}(2019)\citenamefont
  {Slobinsky}, \citenamefont {Pili},\ and\ \citenamefont {Borzi}}]{Slobinsky}%
  \BibitemOpen
  \bibfield  {author} {\bibinfo {author} {\bibfnamefont {D.}~\bibnamefont
  {Slobinsky}}, \bibinfo {author} {\bibfnamefont {L.}~\bibnamefont {Pili}}, \
  and\ \bibinfo {author} {\bibfnamefont {R.A.}\ \bibnamefont {Borzi}},\
  }\bibfield  {title} {\enquote {\bibinfo {title} {The polarized monopole
  liquid: a $\textup{C}$oulomb phase in a fluid of magnetic charges},}\
  }\href@noop {} {\bibfield  {journal} {\bibinfo  {journal} {arXiv:1903.11493}\
  } (\bibinfo {year} {2019})}\BibitemShut {NoStop}%
\bibitem [{\citenamefont {Gray}(2019)}]{Gray}%
  \BibitemOpen
  \bibfield  {author} {\bibinfo {author} {\bibfnamefont {C.}~\bibnamefont
  {Gray}},\ }\emph {\bibinfo {title} {{Field correlations in $\textup{C}$oulomb
  gases}}},\ \href@noop {} {Ph.D. thesis},\ \bibinfo  {school} {University
  College London, UCL}, \bibinfo {address} {United Kingdom} (\bibinfo {year}
  {2019})\BibitemShut {NoStop}%
\end{thebibliography}
\end{document}